\documentclass{gGAF2e}

\usepackage{multiequations}

\newcommand*{\bme}{\begin{multiequations}}
\newcommand*{\eme}{\end{multiequations}}
\newcommand*{\bse}{\begin{subequations}}
\newcommand*{\ese}{\end{subequations}}
\newcommand*{\be}{\begin{equation}}
\newcommand*{\ee}{\end{equation}}
\newcommand*{\se}{\singleequation}
\newcommand*{\de}{\doubleequation}
\newcommand*{\te}{\tripleequation}

\renewcommand{\vec}[1]{\bm #1}
\newcommand{\tensor}[1]{\bm #1}
\newcommand{\nond}[1]{\hat{#1}}
\renewcommand*{\Gamma}{{\varGamma}}
\renewcommand*{\Omega}{{\varOmega}}
\renewcommand*{\Pi}{{\varPi}}

\begin{document}

\jvol{00} \jnum{00} \jyear{2017}

\markboth{J.~Verhoeven and G.A.~Glatzmaier}{Geophysical and Astrophysical Fluid Dynamics}

\title{Validity of sound-proof approaches in rapidly-rotating compressible convection: Marginal stability versus turbulence}

\author{Jan Verhoeven$^{\ast}$\thanks{$^\ast$Corresponding author. Email: jverhoev@ucsc.edu
\vspace{6pt}} and Gary A.~Glatzmaier\\\vspace{6pt}  Earth and Planetary Sciences Department, University of California Santa Cruz, CA, USA\\\vspace{6pt}\received{v4.4 released October 2012} }

\maketitle

\begin{abstract}
The validity of the anelastic approximation has recently been questioned in the regime of rapidly-rotating compressible convection in low Prandtl number fluids \citep{Calkins2015}. Given the broad usage and the high computational efficiency of sound-proof approaches in this astrophysically relevant regime, this paper clarifies the conditions for a safe application. The potential of the alternative pseudo-incompressible approximation is investigated, which in contrast to the anelastic approximation is shown to never break down for predicting the point of marginal stability. Its accuracy, however, decreases close to the parameters corresponding to the failure of the anelastic approach, which is shown to occur when the sound-crossing time of the domain exceeds a rotation time scale, i.e. for rotational Mach numbers greater than one. Concerning the supercritical case, which is naturally characterised by smaller rotational Mach numbers, we find that the anelastic approximation does not show unphysical behaviour. Growth rates computed with the linearised anelastic equations converge toward the corresponding fully compressible values as the Rayleigh number increases.  Likewise, our fully nonlinear turbulent simulations, produced with our fully compressible and anelastic models and carried out in a highly supercritical, rotating, compressible, low Prandtl number regime show good agreement. However, this nonlinear test example is for only a moderately low convective Rossby number of $0.14$.


\end{abstract}

\section{Introduction}
\label{introduction}

The interiors of stellar and many planetary bodies can be characterised as compressible rapidly-rotating fluids that have small viscous compared to thermal diffusivities, i.e., Prandtl numbers much less than unity.
These astrophysical fluid dynamical systems can be modeled with the fully compressible equations that follow from first principles of physics.
Their generality comprises a broad range of temporal and spacial scales corresponding to physical phenomena reaching from sound waves over buoyancy induced flows up to atmospheric jets.
Accounting for each of these processes, however, implies high computational costs for numerical simulations.
In order to more adequately understand the structure and evolution of planetary and astrophysical objects, it is often favourable to channel all available resources into only the relevant physical phenomena, which has motivated the development of reduced forms of the fully compressible equations.
For the purpose of modeling planetary and stellar convection, it is believed that dynamically unimportant sound waves can  safely be neglected, which, due to the reduced spectrum of time-scales that needs to be resolved, effectively decreases the necessary computational effort.

Several so-called sound-proof approaches retaining the major compressibility effects are in use for modeling these kind of systems. The most widely employed ones are the anelastic approximation (e.g. \citealp{Batchelor1953, Ogura1962, Gough1969, Gilman1981, Braginsky1995,Lantz1999,Jones2009b, Glatzmaier2014, Verhoeven2015}), the pseudo-incompressible approximation \citep{Durran1989,Durran2008,Klein2012,Vasil2013,Wood2016}, and the closely related low Mach number approach \citep{Majda1985,Bell2004,Almgren2006a,Almgren2006b,Almgren2014}.
Two necessary conditions for all these approaches to be valid are that the Mach number, which is defined as the ratio of the fluid velocity to the local sound speed, is much smaller than unity and that pressure perturbations are small relative to the depth-dependent background pressure.
For the pseudo-incompressible and low Mach number approaches this is also sufficient, but for the anelastic approximation all thermodynamic perturbations from a depth-dependent background state must also be small.

Only few studies are available that verify these theoretical predictions quantitatively for thermal convection, with most of them confirming a good agreement between anelastic and fully compressible computations and paying less attention to the pseudo-incompressible approach.
Whereas \citet{Berkoff2010} addressed linear magneto-convection, \citet{Lecoanet2014} investigated differences between thermal conductivity and large-eddy entropy diffusion \citep{Glatzmaier1984} and additionally considered the pseudo-incompressible case. This was followed by \citet{Calkins2014} verifying the accuracy of the anelastic approximation at the point of marginal stability under the influence of rotation in Prandtl number unity fluids.
However, in a follow-up study \citet{Calkins2015} showed that the linear anelastic equations fail to produce physically meaningful results for marginally stable, rapidly-rotating, low Prandtl number systems. Against their theoretical expectations, they report that this breakdown of the anelastic approximation is caused by the temporal derivative of the density perturbation becoming an important component in the continuity equation. Furthermore, they suspect related problems in the astrophysically relevant nonlinear turbulence regime.
So far, \citet{Verhoeven2015} carried out the only systematic one-to-one comparison for fully compressible and anelastic turbulent convection, which again proved the functionality of the anelastic equations. They, however, did not consider the problematic regime of rapid-rotation and low Prandtl number.

Although there has been no such study that proves the breakdown of the anelastic approximation for the nonlinear case, the findings of \citet{Calkins2015} are alarming, as they implicitly question the validity of works that investigate astrophysical objects by means of the anelastic equations.
There are two possible ways to deal with this problem: Firstly, alternative sound-proof approaches must be studied concerning their applicability at the point of marginal stability. Secondly, the possible breakdown of the anelastic approximation in the astrophysically more relevant turbulence regime must be either confirmed or disproved.
Here we extend the work of \citet{Calkins2015} on rapidly-rotating, low Prandtl number compressible convection and present a study on the accuracy of the linear pseudo-incompressible equations at the point of marginal stability. This approach is promising as it retains parts of the problematic temporal derivative of the perturbational density term and thus may provide a more accurate alternative to the anelastic approximation. Furthermore, we present the first one-to-one comparison between fully compressible and anelastic numerical simulations of rapidly-rotating convection in the fully nonlinear turbulent regime at low Prandtl number.

The following questions are addressed:
\begin{itemize}
\item How do the anelastic and the pseudo-incompressible approaches vary with respect to their accuracy in predicting the point of marginal stability for rapidly-rotating compressible convection in low Prandtl number fluids?
\item Can we identify the physical process constraining their accuracy, which causes the problems found by \citet{Calkins2015}?
\item Is it safe to apply the anelastic approximation to the astrophysically relevant fully nonlinear turbulent regime of rapidly-rotating compressible convection at low Prandtl number?
\end{itemize}

The remainder of this paper is organised as follows. In section \ref{model}, we start with defining our idealised model and discussing the differences among the anelastic, pseudo-incompressible and fully compressible approaches. Then in section \ref{results} we describe our numerical results concerning the marginal point of stability, supercritical linear convection and the fully nonlinear turbulence regime. Finally, general conclusions are drawn in section \ref{conclusions}.

\section{Model}
\label{model}

In this section the fully compressible, pseudo-incompressible and anelastic linear equations are discussed along with the numerical approach to solve them.

\subsection{Governing equations}
\label{equations}

Linear compressible convection within a Newtonian ideal gas is investigated in a plane layer geometry rotating with angular velocity $\vec\Omega=\Omega\nond{\vec z}$. The rotation axis is aligned with the unit vector $\nond{\vec z}$ and antiparallel to the constant gravity $\vec g=-g\nond{\vec z}$. The fluid is characterised by constant specific heat capacities at fixed volume and pressure, $c_v$ and $c_p$. The dynamic viscosity and thermal conductivity,
\bme
\label{viscosity-conductivity}
\be
\mu \, =\,\rho\nu \,, \hskip 20mm
k_t \, =\,c_p\rho\kappa\,,
\ee
\eme
are taken as fixed functions of $z$ and relate to the kinematic quantities $\nu$ and $\kappa$, which are the viscous diffusivity and the thermal diffusivity, respectively.

The linear approximation to the governing equations for fully compressible convection describing the temporal evolution of density $\rho$, fluid velocity $\vec v$ and entropy $s$ within a reference frame rotating at angular velocity $\vec \Omega$ are
\begin{subequations}
\begin{align}
\label{continuity}
\upartial_t \rho_1 \, =\, &\, -\, \vec\nabla{\bm \cdot}\left(\rho_0 \vec v\right), \\
\label{momentum}
\rho_0\upartial_t \vec v\, = \,&\, -\, \vec\nabla \left(p_0 + p_1\right) - \left(\rho_0 + \rho_1\right) g \nond{\vec z} + \vec\nabla{\bm\cdot}\tensor\Pi - 2\Omega\rho_0\nond{\vec z}\times\vec v\,, \\
\label{heat}
\rho_0 T_0 \upartial_t s_1 \,=\, &\, -\, v_z \rho_0 T_0 \upartial_z s_0 + \vec\nabla\left[ k_t \vec\nabla \left(T_0 + T_1\right)\right] + S\,,
\end{align}
\end{subequations}
with $t$, $p$ and $T$ specifying time, pressure and temperature, and the indices $0$ and $1$ denoting the z-dependent background state and its time and 3-D space dependent perturbation, respectively. $\Pi_{ij}=\mu\left(\upartial_{j} v_{i} + \upartial_{i} v_{j} - 2/3(\vec\nabla{\bm \cdot}\vec v) \delta_{ij}\right)$ is the viscous stress tensor for a Newtonian fluid and $S$ denotes a heat source or sink. Note that we have neglected centrifugal forces in (\ref{momentum})
that express the conservation of mass, momentum and energy. They can be closed with an equation of state, i.e. the ideal gas law
\bme
\label{state_bg-state}
\be
p_0 \,=\,(c_p-c_v)\rho_0 T_0\,,
\hskip 30mm
\frac{p_1}{p_0}\,=\,\frac{\rho_1}{\rho_0} + \frac{T_1}{T_0}\,,\qquad\quad
\ee
\eme
and a thermodynamic expression relating entropy, temperature and pressure,
\bme
\label{therm_relation_bg-therm_relation}
\be
\frac{\upartial_z s_0}{c_p}\, =\, \frac{\upartial_z T_0}{T_0} - \frac{c_p-c_v}{c_p}\frac{\upartial_z p_0}{p_0}\,,
\hskip 20mm
\frac{s_1}{c_p} \, =\, \frac{T_1}{T_0} - \frac{c_p-c_v}{c_p}\frac{p_1}{p_0}\,.\quad
\ee
\eme
The background state is chosen to satisfy hydrostatic and thermal equilibrium
\bme
\label{hydrostaticity-therm_equilibrium}
\be
\upartial_z p_0\, =\, -\, \rho_0 g\,,
\hskip 25mm
\upartial_z \left(k_t \upartial_{z} T_0\right) \, =\, -\,S\,,\qquad
\ee
\eme
which results in these background terms dropping out of equations (\ref{momentum}) and (\ref{heat}), respectively. Furthermore, the background temperature gradient
\begin{align}
\label{temp_gradient_bg}
\upartial_z T_0 = -\left(\frac{g}{c_p}+\frac{\Delta T}{d}\right)
\end{align}
is assumed constant in $z$ and defined by the sum of the adiabatic temperature gradient $-g/c_p$ and the ratio of superadiabatic temperature drop, $\Delta T$, prescribed by the boundary conditions and the domain depth $d$. The background state results from solving equations (\ref{state_bg-state}a),
(\ref{therm_relation_bg-therm_relation}a),
(\ref{hydrostaticity-therm_equilibrium}a)
and (\ref{temp_gradient_bg}). Therefore, for a constant thermal conductivity, $S$ vanishes; but, for a constant thermal diffusivity, $S$ is negative, i.e., a heat sink.

\subsection{Non-dimensionalisation and parameters}
\label{non-dim_equations}

The background state can be non-dimensionalised by using the bottom temperature $T_r$, bottom density $\rho_r$, bottom pressure $p_r=(c_p-c_v)\rho_r T_r$ and $c_p$ for entropy. The superadiabatic temperature difference $\Delta T$ prescribed by the boundary conditions of the system is chosen to be the scale for the temperature perturbations. As temperature, density and entropy perturbations are usually assumed to be closely correlated in low Mach number flows (see e.g. \citealp{Clayton1968}), the density and entropy perturbations are scaled correspondingly with $\Delta \rho=\rho_r \Delta T / T_r$ and $\Delta s=c_p \Delta T / T_r$. The domain depth $d$ is used as reference length scale. The velocity is non-dimensionalised with a convective free-fall velocity $v_r=\sqrt{\Delta\rho g d / \rho_{r}}$ and correspondingly time is scaled with the free-fall time $t_r=d/v_r=\sqrt{\rho_r d / (\Delta\rho g)}$. The pressure perturbation scale $\Delta p = \rho_r v_r^2=\Delta\rho g d$ is inferred from the fact that pressure extracts kinetic energy from the vertical flows to drive horizontal motions. The scales for viscous and thermal diffusivities $\nu_r$ and $\kappa_r$ are defined as their respective values at the bottom boundary of the domain.

When applying these scales, the following non-dimensional parameters emerge.
The dissipation number,
\begin{align}
  \label{diss}
  D\,=\,{g d}\big/({c_p T_r})\,,
\end{align}
defines the absolute value of the non-dimensional global adiabatic temperature gradient. With $-(D+\epsilon)$ being the total dimensionless temperature gradient, see (\ref{T_0-rho_0-p_0-s_0}a),
the superadiabaticity of the system is given by
\begin{align}
  \label{epsilon}
    \epsilon\,=\,{\Delta T}\big/{T_r}\,.
\end{align}
The polytropic index
\begin{align}
\label{polyindex}
n\,=\,\frac{\gamma}{\gamma-1}\frac{D}{D+\epsilon}\,-\,1\,,
\end{align}
with the ratio of specific heats
\begin{align}
  \label{gamma}
  \gamma\,=\,{c_p}\big/{c_v}\,=\,{5}/{3}
\end{align}
being chosen to represent a monatomic ideal gas, is an alternative (dependent) parameter for the superadiabaticity $\epsilon$, with a superadiabatic polytropic index satisfying $n<1.5$. The adiabatic polytropic index for an ideal monatomic ideal gas (with $\epsilon=0$) consistently results in
\begin{align}
\label{polyindex_ad}
n_{ad}=\frac{1}{\gamma-1}=1.5.
\end{align}
The ratio of superadiabatic to adiabatic temperature gradient $\epsilon/D$ often given by 1-D solar models\footnote{\citet{Christensen1996} for example specify values for $\epsilon/D$ of ${\mathrm O}(10^{-4})$ in the bulk of the solar convection zone.} follow from the derivatives of equations (\ref{T_0-rho_0-p_0-s_0}a,b).
\begin{align}
\frac{\epsilon}{D}=\frac{n_{ad}-n}{n+1}.
\end{align}
Another useful and widely used parameter is the number of density scale heights
\begin{align}
  \label{dsh}
  N_\rho \,= \,\ln\bigl({\rho_{\mathrm {bot}}}\big/{\rho_{\mathrm {top}}}\bigr)\,=\,\,-\,n\,\ln\left[1-(D+\epsilon)\right]
\end{align}
specifying the system's density contrast $\rho_{\mathrm {bot}}/\rho_{\mathrm {top}}$ between bottom and top boundary that can be used instead of the Dissipation number $D$.
The Prandtl number,
\begin{align}
  \label{pr}
  Pr\,=\,{\nu_r}\big/{\kappa_r}\,,
\end{align}
is the ratio of momentum to thermal diffusivity.
The Rayleigh number,
\begin{align}
  \label{ra}
  Ra\,=\,{g d^3 \Delta T}\big/\bigl({\kappa_r \nu_r T_r}\bigr)\,,
\end{align}
controls the vigour of convection with large $g$, $d$, and $\epsilon=\Delta T / T_r$ enhancing and large diffusivities $\nu$ and $\kappa$ reducing the convective vigour. More formally $Ra$ is the ratio of the product of the diffusive timescales $d^2/\kappa_r ~ d^2/\nu_r$ to the square of the free-fall timescale $t_r^2$. The Ekman number,
\begin{align}
  \label{ek}
  Ek\,=\,{\nu_r}\big/\bigl({2\Omega d^2}\bigr)\,,
\end{align}
is the ratio of the rotation time scale to the viscous diffusion time scale.

Note that all non-dimensional parameters given in this subsection are defined at the bottom boundary and may vary strongly over the whole domain.

\subsection{Non-dimensional equations}

The background state results in
\bme
\label{T_0-rho_0-p_0-s_0}
\begin{align}
\nond{T}_0(\nond{z})\,= \,&\, 1-(D+\epsilon)\nond{z}, &
\nond{\rho}_0(\nond{z})\,=\, & \,\left[1-(D+\epsilon)\nond{z}\right]^n, \\
\nond{p}_0(\nond{z})\,=\, &\, \left[1-(D+\epsilon)\nond{z}\right]^{n+1}, &
\nond{\upartial}_z\nond{s}_0(\nond{z})\, =\, & \,-\,{\epsilon}\big/{\nond{T}_0}\,,
\end{align}
\eme
and the governing non-dimensional equations read 
\bme
\label{continuity_momentum_heat_state_therm-relation_nondim1}
\se
\begin{align}
\frac{\epsilon D}{\gamma - 1}\frac{1}{\nond{T}_0}\nond{\upartial}_t \nond{p}_1 - \epsilon \nond{\rho}_0 \nond{\upartial}_t \nond{s}_1 \,=\,&\, - \nond{\vec\nabla}{\bm \cdot}\left(\nond{\rho}_0 \nond{\vec v}\right), \\[-0.4em]
\nond{\rho}_0\nond{\upartial}_t \nond{\vec v} \,=\,& \,-  \nond{\vec\nabla} \nond{p}_1 -  \nond{\rho}_1 \nond{\vec z} + \sqrt{\frac{Pr}{Ra}} \nond{\vec\nabla}{\bm \cdot}\nond{\tensor\Pi} - \frac{1}{Ek}\sqrt{\frac{Pr}{Ra}}\nond{\rho}_0\nond{\vec z}\times\nond{\vec v}\,, \qquad\qquad\\
\nond{\rho}_0 \nond{T}_0 \nond{\upartial}_t \nond{s}_1 \,=\, & \,\nond{\rho}_0 \nond{v}_z + \frac{1}{\sqrt{Pr Ra}}\nond{\vec\nabla}\bigl(\nond{k}_t \nond{\vec\nabla} \nond{T}_1\bigr),
\end{align}
\vskip -5mm
\be
\de
\nond{\rho}_1 \,= \,\frac{ D}{\gamma - 1}\frac{1}{\nond{T}_0} \nond{p}_1 -  \nond{\rho}_0 \nond{s}_1, \hskip 20mm
\nond{s}_1 \,=\,  \frac{\nond{T}_1}{\nond{T}_0} -  D \frac{\nond{p}_1}{\nond{p}_0}\,,\qquad\qquad
\ee
\eme
with the hat $\nond{}$ denoting non-dimensional quantities, e.g. $T_0=T_r \nond{T}_0$.

\subsection{Differences in the anelastic, pseudo-incompressible and fully compressible equations}

The density perturbation term in the continuity equation (\ref{continuity_momentum_heat_state_therm-relation_nondim1}a)
has been expressed in terms of pressure and entropy by using (\ref{continuity_momentum_heat_state_therm-relation_nondim1}d)
in order to illustrate the approximations carried out in the anelastic and pseudo-incompressible equations in the following.

The anelastic approximation neglects the time derivative of density $\epsilon \nond{\upartial}_t\nond{\rho}_1 = \epsilon D/(\gamma - 1)/\nond{T}_0 \nond{\upartial}_t\nond{p}_1 - \epsilon \nond{\rho}_0 \nond{\upartial}_t\nond{s}_1$ in the continuity equation (\ref{continuity_momentum_heat_state_therm-relation_nondim1}a),
which is reasonable as long as all perturbations remain small with $\epsilon \ll 1$. The pseudo-incompressible approximation is less restrictive and only neglects the time derivative of the perturbational pressure term. This approach is justified for low Mach number flows but allows for temperature, density and entropy perturbations to be large\footnote{Note firstly that the prefactor in the perturbational pressure term equals the squared Mach number (see Appendix \ref{appendix_mach_reg}) and secondly that $-\rho_0/c_p\upartial_t s_1$, which is the dimensional form of the $- \epsilon \nond{\rho}_0 \nond{\upartial}_t \nond{s}_1$ term in the continuity equation (\ref{continuity_momentum_heat_state_therm-relation_nondim1}a), equals the temporal derivative of the pseudo-incompressible density $\upartial_t \rho^*$ as first defined in \cite{Durran1989}.}.
In order to check for the validity of both sound-proof approaches the relative magnitudes of the terms being neglected in each approximation, i.e. the perturbational density and pressure terms in the continuity equation
\begin{subequations}
\begin{align}
\label{dens_part}
\rho_{\mathrm {cont}}\, =\, & \,\Biggl(\frac{\bigl| \epsilon\nond{\upartial}_t \nond{\rho}_1\bigr|}{{\mathrm {max}}\bigl[\,\bigl|\nond{\upartial}_x (\nond{\rho}_0 \nond{v}_x)\bigr|,\,\bigl|\nond{\upartial}_y (\nond{\rho}_0 \nond{v}_y)\bigr|,\,\bigl|\nond{\upartial}_z (\nond{\rho}_0 \nond{v}_z)\bigr|\,\bigr]}\Biggr)^{\!\!{\mathrm {top}}}, \\
\label{pres_part}
p_{\mathrm {cont}}\, =\, &\, \Biggl(\frac{\epsilon D}{\gamma - 1}\,\frac{1}{\nond{T}_0} \,\,\frac{\bigl| \nond{\upartial}_t \nond{p}_1\bigr|}{{\mathrm {max}}\bigl[\,\bigl|\nond{\upartial}_x (\nond{\rho}_0 \nond{v}_x)\bigr|,\,\bigl|\nond{\upartial}_y (\nond{\rho}_0 \nond{v}_y)\bigr|,\,\bigl|\nond{\upartial}_z (\nond{\rho}_0 \nond{v}_z)\bigr|\,\bigr]}\Biggr)^{\!\!{\mathrm {top}}},
\end{align}
\end{subequations}
will be quantified for various parameters in this paper. We have chosen to focus on the values at the top boundary as the sound speed is the lowest at this location.

For simplicity we will assume $\epsilon\rightarrow 0\Longleftrightarrow n\rightarrow 1.5$ for each anelastic simulation carried out making the $\epsilon$ parameter obsolete. This yields the same equations as when assuming a small but finite $\epsilon$ and neglecting all terms involving $\epsilon$ in (\ref{T_0-rho_0-p_0-s_0}a-d) and
(\ref{continuity_momentum_heat_state_therm-relation_nondim1}a)
resulting in an adiabatic background state, which is perturbed by a constant superadiabatic temperature drop prescribed by the boundary conditions. Similar approaches are common practice when applying the anelastic approximation for nonlinear turbulent convection simulations \citep[see, e.g.][]{Gastine2014,Heimpel2015} and do not increase the error of the anelastic approximation \citep{Lantz1999}. Concerning the pseudo-incompressible approach, however, the parameter $\epsilon$ is required, as it still appears in the continuity equation (\ref{continuity_momentum_heat_state_therm-relation_nondim1}a)
in the entropy term.

\subsection{Numerical approach for the linear stability problem}
\label{numerics}

In order to solve the linear equations (\ref{continuity_momentum_heat_state_therm-relation_nondim1}a-e)
numerically, each variable is represented by the typical normal mode ansatz, e.g., $\nond{T}_1=\nond{T}(\nond{z})\exp\bigl[\nond{r}\nond{t}+{\mathrm i}\bigl(\nond{\omega} \nond{t} + \nond{k}_x \nond{x} + \nond{k}_y \nond{y}\bigr)\bigr]$. Here, $\nond{r}$ denotes the growth rate, $\nond{\omega}$ is the oscillation frequency and $\nond{k}_x$ and $\nond{k}_y$ are the horizontal wavenumbers with $\nond{k}=\sqrt{\nond{k}_x^2+\nond{k}_y^2}$ (not to be confused with the thermal conductivity $\nond{k}_t$).
The equations, as they are solved numerically, are given in Appendix \ref{appendix_numerics}, with the critical frequency $\nond{\omega}_c$ and the critical Rayleigh number $Ra_c$ being eigenvalues that depend on the critical wavenumber $\nond{k}_c$, which is determined by using a nested intervals scheme for $\nond{r}=0$.
Growth rates $\nond{r}$ can be determined by prescribing the Rayleigh number $Ra$ and the wavenumber $\nond{k}$. A positive $\nond{r}$ with a non-zero $\nond{\omega}$ means that the onset of convection is a temporal oscillation of all variables with their mean amplitudes increasing exponentially in time, whereas the instability is non-oscillatory for $\nond{\omega}=0$.
As the solutions corresponding to critical modes $\nond{k}_c$ are independent of the individual choice of $\nond{k}_x$ and $\nond{k}_y$ and manifest themselves in two-dimensional convection rolls, we restrict our linear analysis to the $k_y=0$ modes for simplicity without losing generality \citep[see][]{Calkins2015}.

The numerical framework is based on a second order accurate Newton-Raphson-Kantorovich (NRK) method for solving eigenvalue problems.
While the computational domain is periodic in horizontal direction, stress free, impermeable and fixed temperature (i.e. $\nond{T}_1=0$) top and bottom boundaries are applied. 

The numerical approaches for solving the nonlinear fully compressible and anelastic equations given in Appendix \ref{appendix_nonlinear_eq} are described in \citet{Verhoeven2014,Verhoeven2015}.

\section{Results}
\label{results}

In this section results from a suite of anelastic, pseudo-incompressible and fully compressible computations are presented in order to test their accuracy and limitations.
All linear computations shown are carried out at fixed Ekman number $Ek=10^{-6}$ and reside in the rapidly-rotating geostrophic limit at low Rossby number, i.e. the Coriolis forces are mostly balanced by the pressure gradient and therefore the results will be essentially the same for computations with smaller Ekman numbers. This was shown by \citet{Calkins2015} for the marginal stability cases and will be checked for the supercritical simulations presented in the following.
We start with studying the point of marginally stable convection, which is followed by investigating the onset of convection in a supercritical setup and eventually finish with the fully nonlinear turbulence regime.

\subsection{Marginal stability}
\label{onset_conv}

\begin{figure}
\centering
\includegraphics[width=0.85\linewidth]{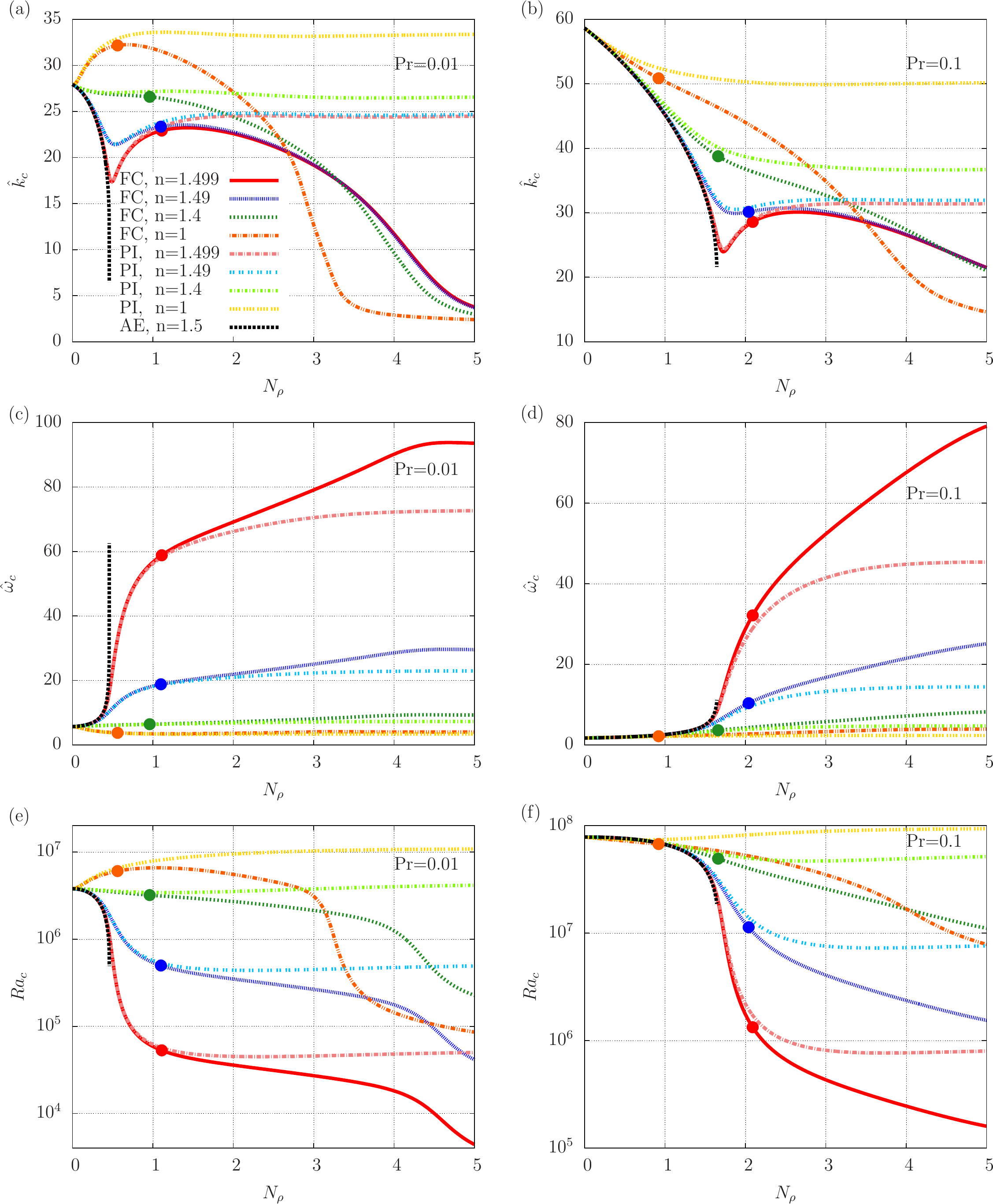}
\caption{Critical wavenumbers $\nond{k}_c$ (a,b), frequencies $\nond{\omega}_c$ (c,d) and Rayleigh numbers $Ra_c$ (e,f) are plotted against the number of density scale heights $N_\rho$ for different polytropic indices $n$ and $Pr=0.01$ (a,c,e) and $Pr=0.1$ (b,d,f). The dynamic viscosity $\mu$ and thermal conductivity $k_t$ are constant and the Ekman number $Ek=10^{-6}$ is fixed for all cases. Results obtained from the pseudo-incompressible equations (PI) are plotted in lighter colours and match the ones provided by the fully compressible equations (FC) in darker colours as long as the relative magnitude of the perturbational pressure term $p_{\mathrm {cont}}$ is negligible in the continuity equation, see (\ref{pres_part}). The transitions to relative magnitudes higher than $0.1$ are marked with circles on the fully compressible curves. As shown by \citet{Calkins2015} the anelastic equations (AE) fail at finite $N_\rho$ (plotted in black). Interestingly, all approaches show the general trend of considerably decreasing critical Rayleigh numbers with decreasing superadiabaticity, i.e., $n\rightarrow 1.5$, at strong density contrasts. (Colour online)}
\label{figure1}
\end{figure}

Although linear marginal modes at the onset of convection are not relevant to turbulent convection in stars and planets, we start with the investigation of this topic. This will pave the way for understanding possible problems in sound-proof approaches for supercritical cases. Accordingly, some of the parameter values considered in this subsection do not correspond to an astrophysically achievable system.

Figure~\ref{figure1} shows plots of critical wavenumbers $\nond{k}_c=\sqrt{\nond{k}_x^2+\nond{k}_y^2}$, frequencies $\nond{\omega}_c$ and Rayleigh numbers $Ra_c$ against the number of density scale heights $0 \le N_\rho \le 5$ for all approaches with various polytropic indices $1\le n \le 1.5$ at low Prandtl numbers $0.01 \le Pr \le 0.1$.
The results from \citet{Calkins2015} were reproduced for the anelastic and the fully compressible equations for these constant dynamic viscosity $\mu$ and thermal conductivity $k_t$ cases.\footnote{Compare figures~(\ref{figure1}a,b,c,d) to \citet{Calkins2015} figures~(3a,c,b,d), respectively. Note that they plot slightly different values than we do. While they plot $\nond{k}_c Ek^{1/3}$, we plot $\nond{k}_c$, which effectively makes our values $100$ times larger for the $Ek=10^{-6}$ cases shown. Moreover they scale their critical frequency $\nond{\omega}_c^{\mathrm{Calkins}}$ with a timescale inferred from the free-fall velocity based on values at the top boundary, see their equation (2.6). We, however, use reference values at the bottom boundary. The relation between our critical frequency and theirs is given by $\nond{\omega}_c=\exp\bigl[(N_\rho)^{1/(2n)}\bigr]\nond{\omega}_c^{\mathrm{Calkins}}$. See also their supplementary material for critical Rayleigh numbers.}
In contrast to the anelastic approximation that produces infinitely small $Ra_c$ and $\nond{k}_c$ at finite $N_\rho$, which \citet{Calkins2015} considered "unphysical", the pseudo-incompressible approach never fails and always produces physically meaningful results with $Ra_c > 0$ and $\nond{k}_c > 0$ in our computations.
The results obtained with the pseudo-incompressible equations, however, deviate from those of the fully compressible equations when the relative magnitude of the perturbational pressure term $p_{\mathrm {cont}}$ (see (\ref{pres_part})) gets larger than ${\mathrm O}(0.1)$. These transitions are marked with circles in figure~\ref{figure1} and generally correspond to lower $N_\rho$ the lower the Prandtl number.
These results suggest that the anelastic and the pseudo-incompressible approximation both should be applied carefully in marginally stable, rapidly-rotating, low Prandtl number convection systems.

Note that in our simplified model the failure of the anelastic approximation generally involves a drastic decrease in critical wavenumbers $k_c$ corresponding to very large aspect ratios $\Gamma$, which are given by the horizontal wavenumber $k_c$ and the depth of the domain $d$ with $\Gamma=2\pi/(k_c d)=2\pi/\nond{k}_c$. Instead of solving for $\Gamma$, as we do here, one may choose to model marginally stable convection in, for example, a particular spherical shell geometry by prescribing the radii of both the inner and outer boundaries, which sets the value of $\Gamma$ and so places a lower limit on the possible values of the horizontal wavenumber $\nond{k}_c$. How this affects the validity of the anelastic approximation for rapidly-rotating, low Prandtl number, marginally stable convection is not clear.  Two previous studies of this linear stability problem for a 3-D spherical shell \citet{Glatzmaier1981b,Glatzmaier1981d,Drew1995} 
agree for cases with $Pr = 1$ and $10$ but disagree for $Pr = 0.1$.
It is worth noting that many anelastic studies of convection in stars and giant planets have modeled the transport of heat by turbulent eddies as a diffusion process based on a turbulent thermal diffusivity and the gradient of entropy, instead of the gradient of temperature.  The argument \citep{Glatzmaier1984} is that the heat transport by eddies, that are too small to numerically resolve in global models but contribute significantly more heat transport than that by radiation or conduction, should be proportional to the local entropy gradient since turbulence tends to maintain a constant entropy.  \citet{Jones2009b} has shown that using entropy diffusive heat flux in an anelastic linear stability model of rotating marginally stable convection cannot have negative critical Rayleigh numbers as \citet{Drew1995} found for temperature diffusive heat flux at $Pr=0.1$.

\begin{figure}
\centering
\includegraphics[width=0.85\linewidth]{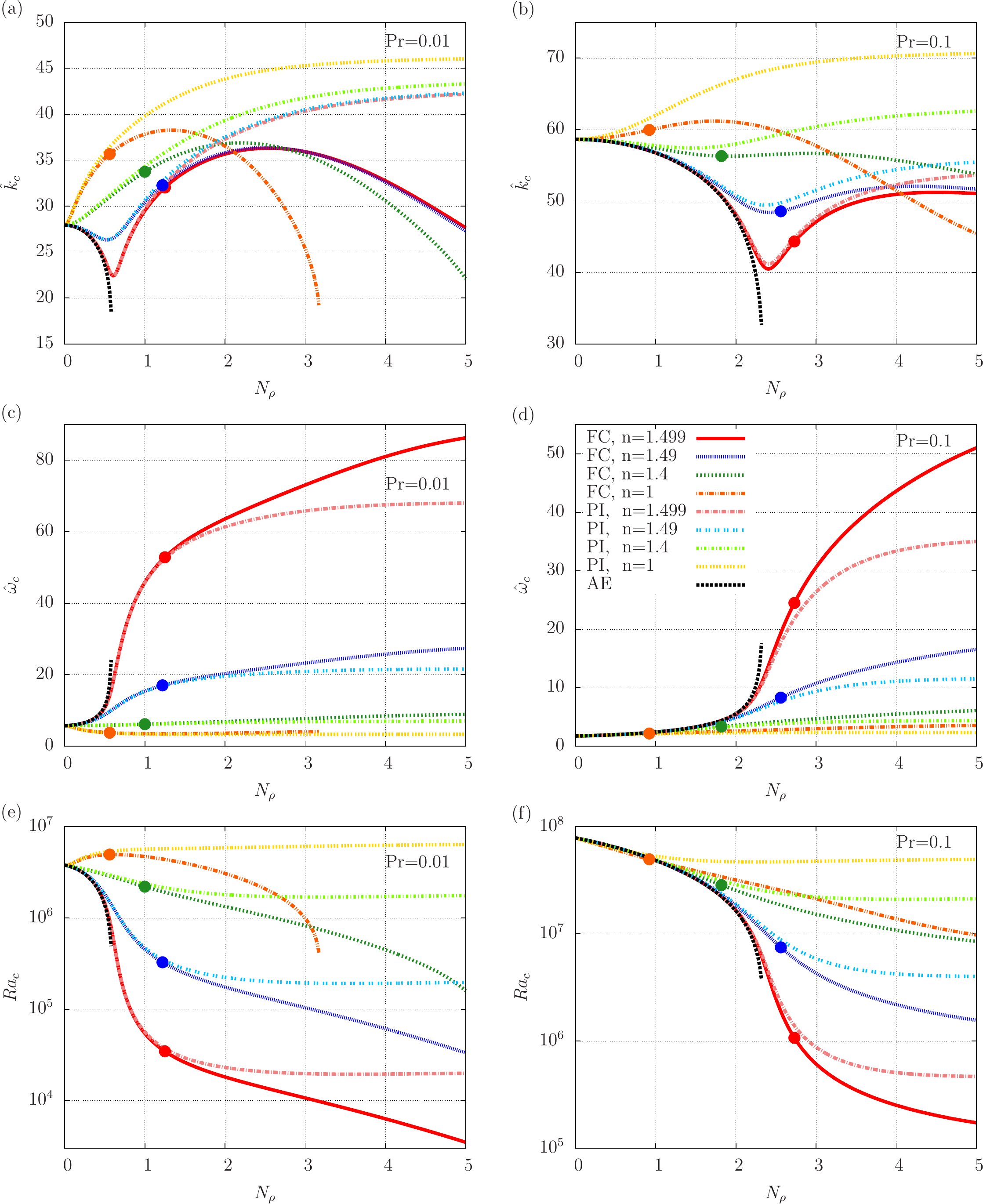}
\caption{The same as in figure~\ref{figure1}, just for constant momentum $\nu$ and thermal diffusivities $\kappa$ rather than constant dynamic viscosity $\mu$ and thermal conductivity $k_t$. The anelastic equations (AE) also fail at finite $N_\rho$ for this case (plotted in black). Interestingly, for $Pr=0.01$ and $n=1$ the fully compressible equations become unstable for infinitely small critical wavenumbers $\nond{k}_c$ at $N_\rho\approx 3.2$ (see orange curve in the left panels). (Colour online)}
\label{figure2}
\end{figure}
The constant $\mu$ and $k_t$ setup and the corresponding parameter range investigated so far, which has been adopted from \citet{Calkins2015}, involves some problems that might explain the observed breakdown of the anelastic approximation and the inaccuracies in the pseudo-incompressible approach: First, the diffusivities $\nu$ and $\kappa$ vary strongly with height for large $N_\rho$ (see (\ref{viscosity-conductivity}a,b))
partly resulting in diffusion velocities close to the sound speed. Second, the phase velocity $\nond{v}_{\mathrm {phase}} = \nond{\omega} / \nond{k}$, which is the velocity of the pattern of the dominant perturbation corresponding to the oscillatory instability through the domain, exceeds the sound speed for specific parameters. Third, for other parameters the typical rotation time $t_{\mathrm {rot}}=1/(2\Omega)$ falls below the sound-crossing time of the the domain $t_{\mathrm {sound}}=d/v_{\mathrm {sound}}$. All of these issues potentially introduce timescales shorter than the free-fall time, which, although $\epsilon$ is small, hinders the justification of the approximations carried out in sound-proof approaches.

In order to exclude the possibility that the accuracy problems of the anelastic and the pseudo-incompressible approach are an artefact due to the very high diffusivities near the top boundary, figure~\ref{figure2} shows the results from a second series of computations using the same parameters as before but for constant diffusivities $\nu$ and $\kappa$ throughout the domain preventing larger diffusion velocities near the top boundary.
Although the plots show individual differences to figure~\ref{figure1}, the overall result is similar.
Alike the anelastic equations that fail to produce physically meaningful outcomes at similar $N_\rho$ as in figure~\ref{figure1}, the pseudo-incompressible approach deviates from the fully compressible method at nearby $N_\rho$ compared to the constant $\mu$ and $k_t$ case.

Interestingly, our fully compressible computation with constant diffusivities, $Pr=0.01$ and $n=1$ yields strongly decreasing critical wavenumbers with $\nond{k}_c \ll 1$ and Rayleigh numbers of ${\mathrm O}(10^3)$ as $N_\rho$ approaches a value of approximately $3.2$.
We speculate that this rather surprising result is caused by the influence of the heat sink that is necessary in order to maintain the background state, see (\ref{hydrostaticity-therm_equilibrium}b).
The investigation of this topic is beyond the scope of this paper and left for future studies.

\begin{figure}
\centering
\includegraphics[width=0.85\linewidth]{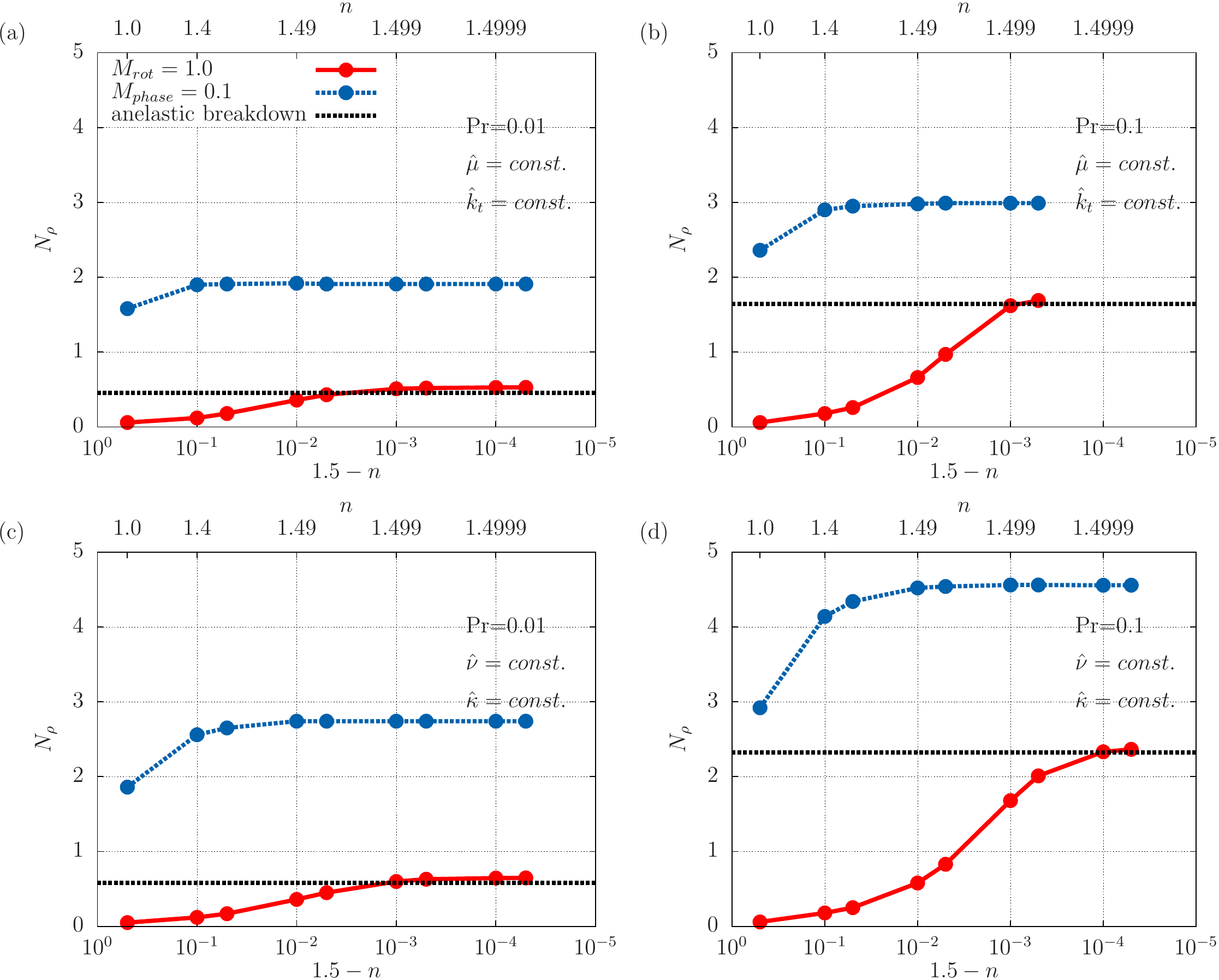}
\caption{The number of density scale heights $N_\rho$, for which the rotational Mach number $M_{\mathrm {rot}}$ reaches a value of $1$ (red) and the phase Mach number $M_{\mathrm {phase}}^{\mathrm {top}}$ equals $0.1$ (blue), is plotted against $n_{ad}-n=1.5-n$ for $Ek=10^{-6}$. It turns out that $N_\rho$ converges against constant values for the anelastic limit $n\rightarrow 1.5$ in all panels representing cases of constant dynamic viscosity and thermal conductivity in the top row (a,b), constant diffusivities at the bottom (c,d), $Pr=0.01$ on the left (a,c) and $Pr=0.1$ on the right (b,d). Additionally, the value for $N_\rho$ corresponding to the breakdown of the anelastic approximation is given by the black line, matching the $N_\rho$ representing $M_{\mathrm {rot}}=1$ in the limit $n\rightarrow 1.5$. (Colour online)}
\label{figure4b}
\end{figure}

According to all our results the anelastic and the pseudo-incompressible approximation both seem to lose accuracy when the perturbational density term and the perturbational pressure term, respectively, have magnitudes that cannot be neglected in the continuity equation anymore, see (\ref{dens_part}-\ref{pres_part}).
In sound-proof approaches it is typically assumed that the magnitude of the neglected perturbational terms relative to the magnitude of the other terms in the continuity equation scale with the square of the Mach number $M^2$, with $M$ being defined as the ratio of a typical fluid velocity $v_{r}$ and the speed of sound $v_{\mathrm {sound}}=\sqrt{c_p(c_p-c_v)T/c_v}$, see Appendix \ref{appendix_mach_reg}.
For linear stability calculations the mean amplitudes of the fluid velocity and thermodynamic perturbations are arbitrary and increase exponentially with time while the background state does not change. Therefore, the regular Mach number $M$ is not a useful diagnostic for linear calculations.  However, the phase velocity $\nond{v}_{\mathrm {phase}} = \nond{\omega} / \nond{k}$ is independent of time in linear computations.
As the sound speed decreases for smaller temperatures, the phase Mach number is the largest at the top boundary where the fluid is cold.
Such a phase top Mach number is derived in the non-dimensionalisation used here in Appendix \ref{appendix_mach_phase} and reads
\begin{align}
\label{M_phase}
M_{\mathrm {phase}}^{\mathrm {top}}&\, = \,M^{\mathrm {top}}\,\frac{\nond{\omega}}{\nond{k}}\, = \,\sqrt{\frac{\epsilon D}{\gamma - 1}} \exp{\left(\frac{N_\rho}{2n}\right)} \frac{\nond{\omega}}{\nond{k}}\,.
\end{align}
Additionally, in order for sound-proof approaches to be valid the rotation time scale must be larger than the sound-crossing time of the domain \citep{Braginsky1995}. A corresponding rotational Mach number can be defined as in appendix \ref{appendix_mach_rot}, resulting in
\begin{align}
\label{M_rot}
M_{\mathrm {rot}} &\, =\,\frac{t_{\mathrm {sound}}^{\mathrm {bot}}}{t_{\mathrm {rot}}}\,= \,\frac{2\Omega d}{v^{\mathrm {bot}}_{\mathrm {sound}}} = \frac{1}{Ek}\sqrt{\frac{\epsilon D Pr}{(\gamma - 1) Ra}}\,.
\end{align}
Please note that for the linear marginal stability case we are considering here, the Rayleigh number in (\ref{M_rot}) must be consistently replaced by $Ra_c$, which is calculated and not a prescribed parameter as it would be for the supercritical onset of convection or a nonlinear simulation. As both, the phase and the rotational Mach numbers, depend on the superadiabaticity $\epsilon$, no finite values can be determined for our anelastic computations with $\epsilon = 0$ directly. Furthermore, in the anelastic limit, $\epsilon\rightarrow 0$, the critical Rayleigh number can become infinitely small, $Ra_c\rightarrow 0$, prohibiting the calculation of $M_{\mathrm {rot}}$. This can be compensated by using fully compressible computations close to the anelastic limit with $\epsilon\ll 1 \Longleftrightarrow 1.5-n \ll 1$. For $Ek=10^{-6}$, $Pr=0.01$ and $Pr=0.1$, and constant $\mu$ and $k_t$ and constant diffusivities $\nu$ and $\kappa$, figure~\ref{figure4b} shows plots of $N_\rho$, for which the afore mentioned Mach numbers exceed a specific limit, against $1.5-n$ being proportional to $\epsilon$ for $\epsilon\ll 1$. The fully compressible simulations are closest to the anelastic limit for small $1.5-n$, for which both Mach number cases show convergence against a constant value of $N_\rho$. Concerning the rotational Mach number case this plateau is caused by the constancy of $\epsilon/Ra_c$ in this parameter range; see figure~\ref{figure1}. Figure~\ref{figure4b} shows that the breakdown of the anelastic approximation (marked by the black line) coincides with the $N_\rho$ corresponding to the rotational Mach number $M_{\mathrm {rot}}$ exceeding $1$ (red). For comparison the values of $N_\rho$ correlating with the phase Mach number $M_{\mathrm {phase}}=0.1$ are displayed in blue. The limit of $0.1$ was chosen, as the limit of $1$ was not exceeded for the computations with small $1.5-n$.

In summary, the accuracy of the anelastic approximation for marginally stable convection is controlled by the rotational Mach number $M_{\mathrm {rot}}$ in our simulations. Our results suggest that \citet{Calkins2015} found the anelastic equations to fail, as the typical rotation time was smaller than the sound-crossing time of the domain in their respective simulations. We speculate that the pseudo-incompressible approximation has a more gentle changeover to being imprecise than the anelastic approximation to failure because the temporal derivative term in the continuity equations compensates for the missing pressure term and allows for numerically stable but nevertheless inaccurate solutions.

\subsection{Supercritical onset of convection}
\label{supercritical_conv}

\begin{figure}
\centering
\includegraphics[width=0.85\linewidth]{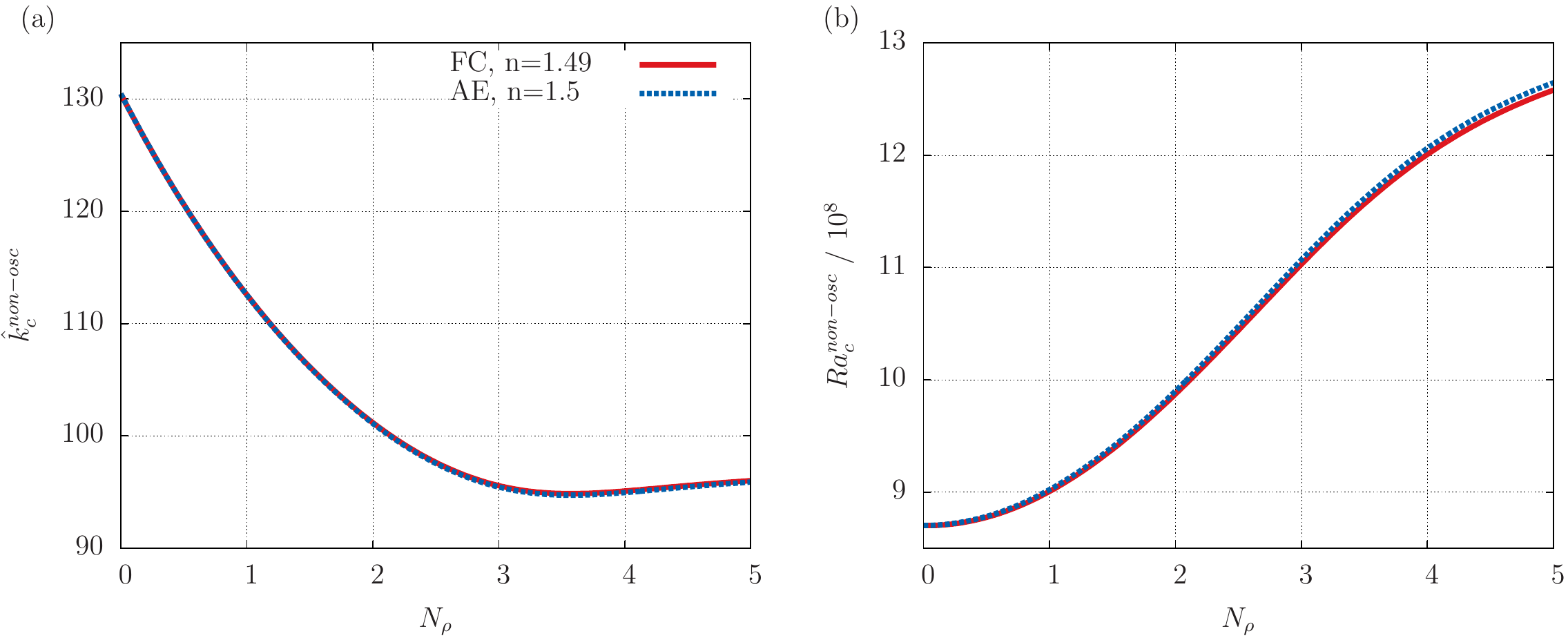}
\caption{Critical wavenumbers $\nond{k}_c^{\mathrm {non-osc}}$ (a) and Rayleigh numbers $Ra_c^{\mathrm {non-osc}}$ (b) for non-oscillatory convection are plotted against the number of density scale heights $N_\rho$ for the fully compressible (red) and the anelastic (blue) case. The results from both approaches closely match and are independent of the Prandtl number $Pr$. The dynamic viscosity $\mu$ and thermal conductivity $k_t$ are constant and the Ekman number $Ek=10^{-6}$ is fixed for both cases. (Colour online)}
\label{figure5}
\end{figure}

\begin{figure}[tb]
\centering
\includegraphics[width=0.85\linewidth]{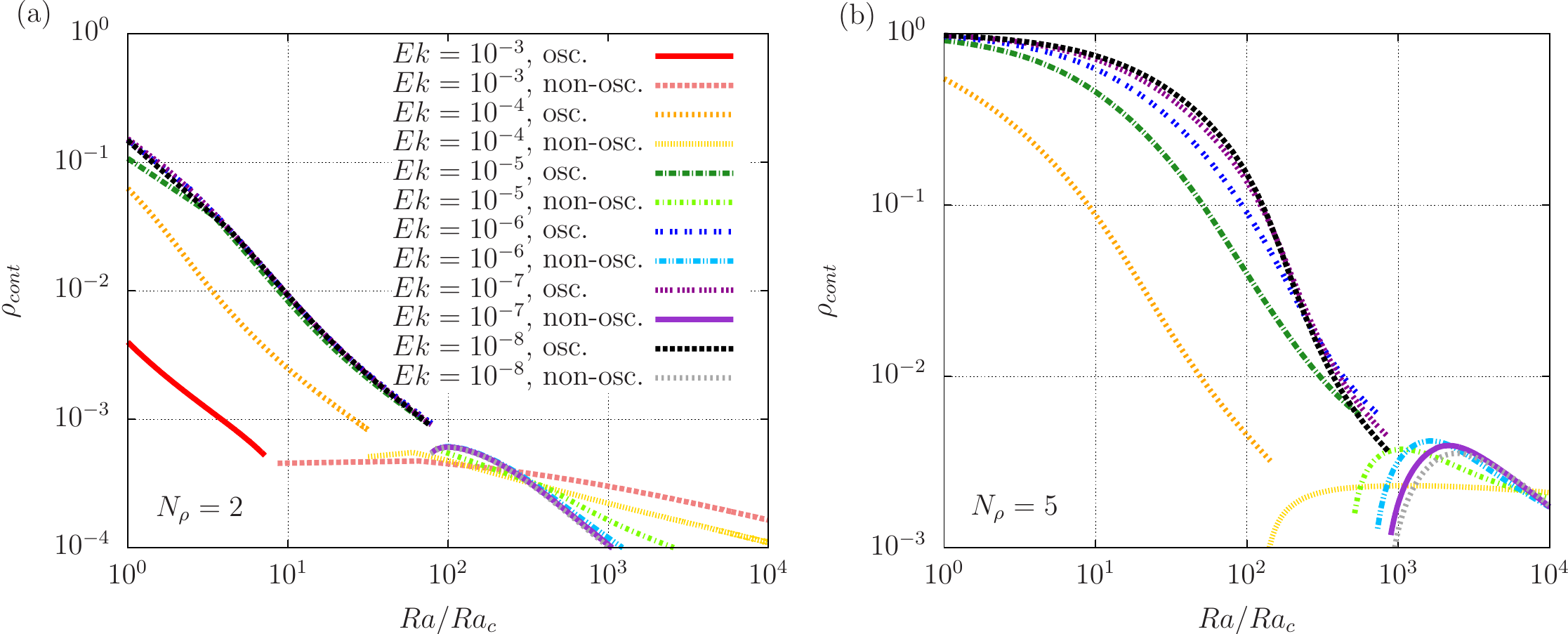}
\caption{The relative magnitudes of the time derivative term in the continuity equation $\rho_{\mathrm {cont}}$ is plotted against the Rayleigh number for oscillatory and non-oscillatory fully compressible linear convection for different Ekman numbers, $n=1.49$, $Pr=0.1$ and $N_\rho = 2$ in panel (a) and $N_\rho=5$ in panel (b). $\rho_{\mathrm {cont}}$ strongly depends on the Rayleigh number and is noncontinuous for $Ra=Ra_c^{\mathrm {non-osc}}$. The time derivative term is negligible with $\rho_{\mathrm {cont}}\ll1$ for all Rayleigh numbers larger than $Ra_c^{\mathrm {non-osc}}$. The asymptotic limit, i.e. geostrophy, is reached for $Ek\le {\mathrm O}(10^{-6})$ for $N_\rho=2$ and $Ek\le {\mathrm O}(10^{-7})$ for $N_\rho=5$. (Colour online)}
\label{figure8a}
\end{figure}

\begin{figure}[tbp!]
\setlength{\unitlength}{\linewidth}
\begin{picture}(1.0,1.05)
\thicklines
\put(0.075,0.0){\includegraphics[width=0.85\linewidth]{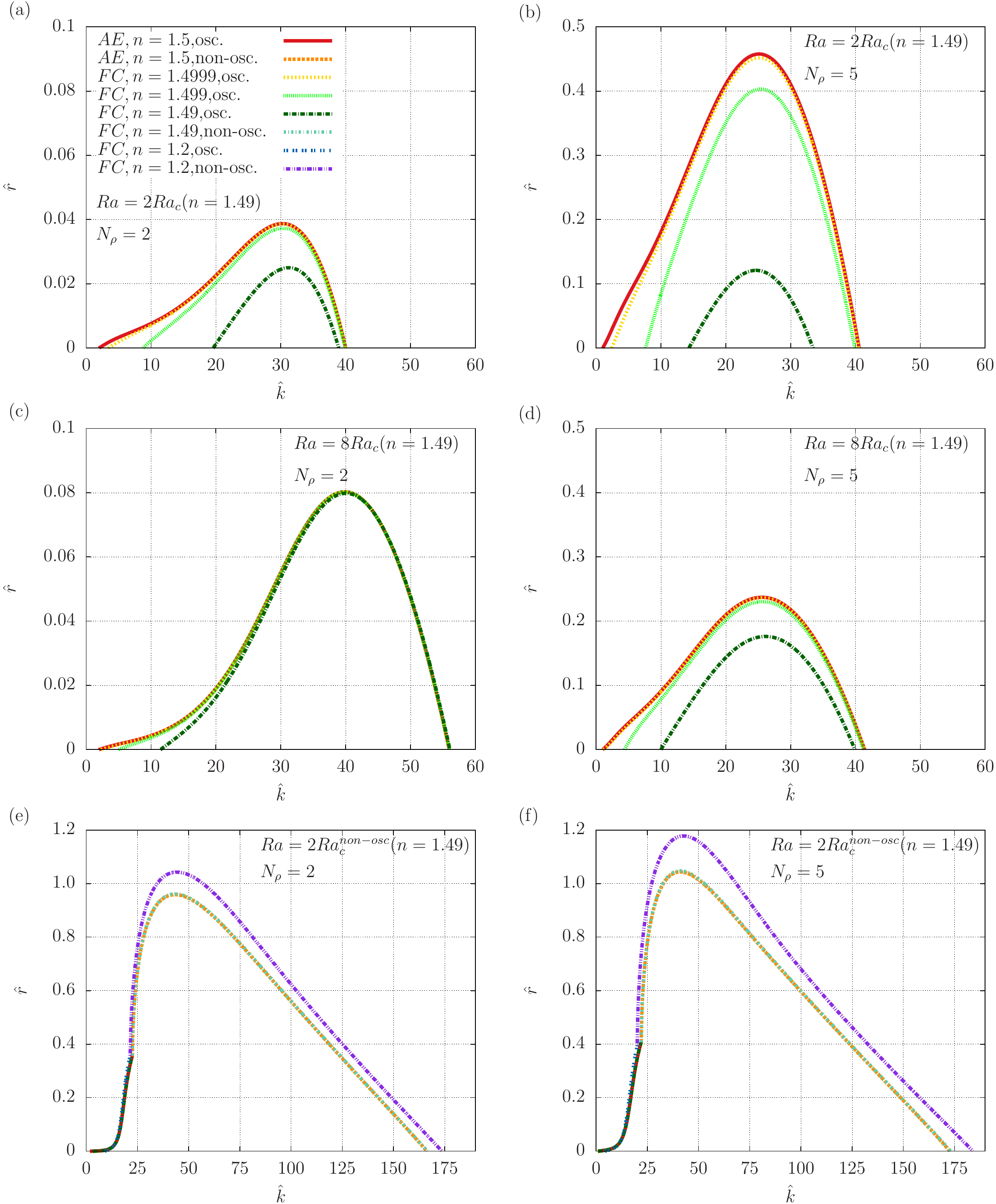}}
\put(0.06,1.04){\vector(1,0){0.3}}
\put(0.06,1.04){\vector(0,-1){0.4}}
\put(0.37,1.03){$N_\rho$}
\put(0.04,0.61){$Ra$}
\end{picture}
\caption{Growth rates $\nond{r}$ are plotted against the wavenumber $\nond{k}$ for different Rayleigh numbers $Ra$, numbers of density scale heights $N_\rho$ and polytropic indices $n$ for $Pr=0.1$. Rayleigh numbers increase from the top to the bottom panels with $Ra=2Ra_c(n=1.49,N_\rho)$ in (a) and (b), $Ra=8Ra_c(n=1.49,N_\rho)$ in (c) and (d) and $Ra=2Ra_c^{\mathrm {non-osc}}(n=1.49,N_\rho)$ in (e) and (f), whereas the number of density scale heights increases from the left to the right panels with $N_\rho=2$ in (a), (c) and (e) and $N_\rho=5$ in (b), (d) and (f). The growth rates of the oscillatory instability associated with different polytropic indices are colour coded for the anelastic limit $n=1.5$ in red, for the fully compressible cases $n=1.4999$ in yellow, $n=1.499$ in light green, $n=1.49$ in dark green and $n=1.2$ in blue. Growth rates of the non-oscillatory instability are plotted in orange for $n=1.5$, in turquoise for $n=1.49$ and in purple for $n=1.2$. While $\nond{r}$ strongly varies between different polytropic indices for large $N_\rho$ and low $Ra$, the match increases for lower $N_\rho$ and larger $Ra$, with the latter being the most constraining parameter. For the $Ra=2Ra_c^{\mathrm {non-osc}}$ cases shown in panels (e) and (f), the growth rates of both kinds of instabilities are indistinguishable for $n\ge 1.49$ and just differ to the ones for $n=1.2$ within a few percent. (Colour online)}
\label{figure8}
\end{figure}
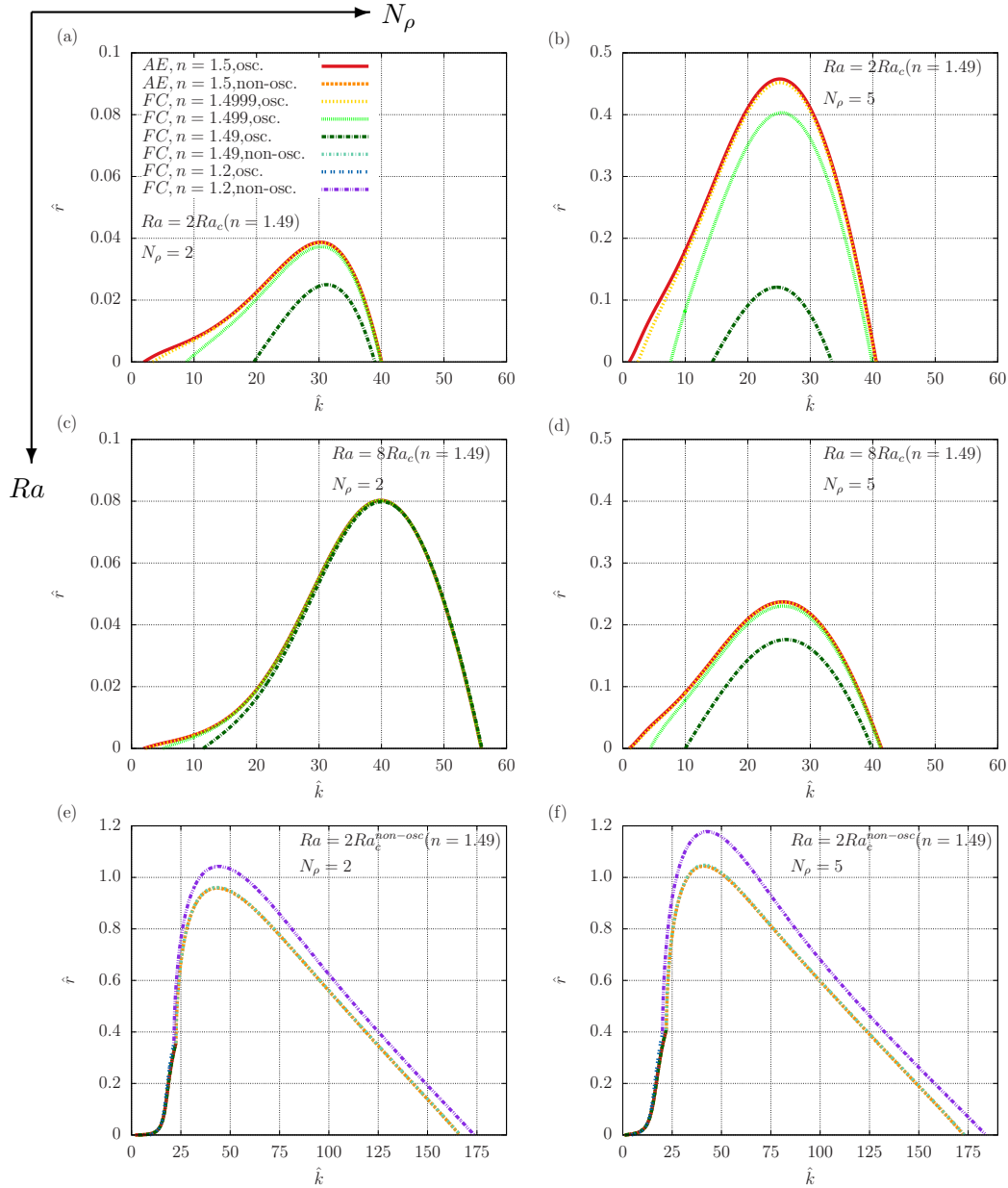

So far, we focused on the marginal point of stability only. Natural systems, however, are often characterised by strongly supercritical Rayleigh numbers, $Ra\gg Ra_c$, which in combination with small superadiabaticity leads to small rotational Mach numbers, see (\ref{M_rot}).
As the supercriticality increases, non-oscillatory (i.e., $\nond{\omega}=0$) convective instabilities emerge. The critical wavenumbers $\nond{k}_c^{\mathrm {non-osc}}$ and Rayleigh numbers $Ra_c^{\mathrm {non-osc}}$ corresponding to this kind of instability do not depend on the Prandtl number and are given in figure~\ref{figure5} for the anelastic ($n=1.5$) and the fully compressible case ($n=1.49$) with $Ek=10^{-6}$ and constant $\mu$ and $k_t$. In contrast to the oscillatory instability, both cases closely match for all $N_\rho$ considered.

The relative magnitudes of the time derivative term in the continuity equation at the top boundary $\rho_{\mathrm {cont}}$, see (\ref{dens_part}), is plotted against the Rayleigh number for oscillatory and non-oscillatory fully compressible linear convection in figure~\ref{figure8a} for $n=1.49$, $Pr=0.1$ and different Ekman numbers and numbers of density scale heights. The outcome of figure~\ref{figure8a} is twofold: First, the results shown do not vary for $Ek\le {\mathrm O}(10^{-6})$ for $N_\rho = 2$ in the left panel, and for $Ek\le {\mathrm O}(10^{-7})$ for $N_\rho = 5$ in the right panel as the geostrophic limit is reached for all corresponding Rayleigh numbers displayed. Second, $\rho_{\mathrm {cont}}$ decreases for increasing $Ra$, with $\rho_{\mathrm {cont}}\ll 1$ for $Ra>Ra_{c}^{\mathrm {non-osc}}$. This strongly suggests the functionality of the anelastic approximation for large Rayleigh numbers for rapidly-rotating convection, although the Prandtl number is low, the density contrast is large and the geostrophic limit is reached.

In order to further ascertain this, the growth rates $\nond{r}$ corresponding to both non-oscillatory and oscillatory instabilities for anelastic and fully compressible convection are displayed in figure~\ref{figure8} for $Ek=10^{-6}$ and $Pr=0.1$.
For given Rayleigh numbers $Ra$, numbers of density scale heights $N_\rho$ and polytropic indices $n$, the growth rates $\nond{r}$ are plotted against the wavenumber $\nond{k}$ .
The anelastic approximation neither fails nor gives rise to spurious growth for any parameters investigated in the supercritical regime. In fact fully compressible growth rates generally converge to the anelastic limit case for $n\rightarrow 1.5$.
While $\nond{r}$ strongly varies between different polytropic indices for large $N_\rho$ and low $Ra$, the match increases for lower $N_\rho$ and larger $Ra$, with the latter seeming to be the most relevant parameter. For moderately high $Ra=2Ra_c^{\mathrm {non-osc}}$ the growth rates of non-oscillatory and oscillatory instabilities are indistinguishable for $n\ge 1.49$ and just differ to the ones for $n=1.2$ within a few percent.

In summary, the problems of the anelastic approximation found by \citet{Calkins2015} only happen close to marginal stability and then only for rotational Mach numbers of order one or larger. They do not occur for Rayleigh numbers more than twice the critical Rayleigh number corresponding to the fully compressible framework.
Given the very high Rayleigh numbers required for nonlinear turbulent convection, this strongly suggests the insignificance of this phenomenon for the turbulence regime relevant to planets and stars.

\subsection{Nonlinear turbulent convection}
\label{turb_conv}

In order to test the functionality of the anelastic approximation in rapidly-rotating turbulent convection in low Prandtl number fluids, a one-to-one comparison with constant $\mu$ and $k_t$ is carried out in the parameter range where \citet{Calkins2015} suspect the breakdown of the anelastic equations.

The employed parameters for the fully nonlinear anelastic simulation are $Pr=0.1$, $Ek=10^{-6}$, $N_\rho=2$ and the Rayleigh number is defined by two times the critical value for non-oscillatory anelastic convection given in figure~\ref{figure5}, which yields $Ra\approx1.98\times10^9$. The choice of these parameters ensures a major role of Coriolis forces with a convective Rossby number \citep{Gilman1977} of $Ro_c=Ek\sqrt{Ra/Pr}\approx 0.14$ and correspond to the geostrophic limit according to the linear results shown in figure~\ref{figure8a}. For the corresponding fully compressible computation we have to additionally provide the superadiabaticity, which is chosen to be $\epsilon=0.1$.
This choice of the superadiabaticity is a trade-off between a low value, which is predicted for planetary and stellar convection zones, to computational feasibility.
We further keep the adiabatic number of density scale heights $N_\rho^{ad}=2$, which yields $N_\rho\approx 2.17$ and $n\approx 1.20$. The aspect ratio $\Gamma=l_x/d=l_y/d$, i.e. the horizontal width divided by the domain height, is inspired by the critical wavenumber associated with the non-oscillatory instability. The choice of $\Gamma\approx 0.39$ ensures that several wavelengths of the critical non-oscillatory ($\sim 6$) and oscillatory instabilities ($\sim 2$) fit into the domain in each horizontal direction, see figures \ref{figure1} and \ref{figure5}.

The anelastic and the fully compressible simulations turn out to be very similar.
\begin{figure}
\centering
\includegraphics[width=0.85\linewidth]{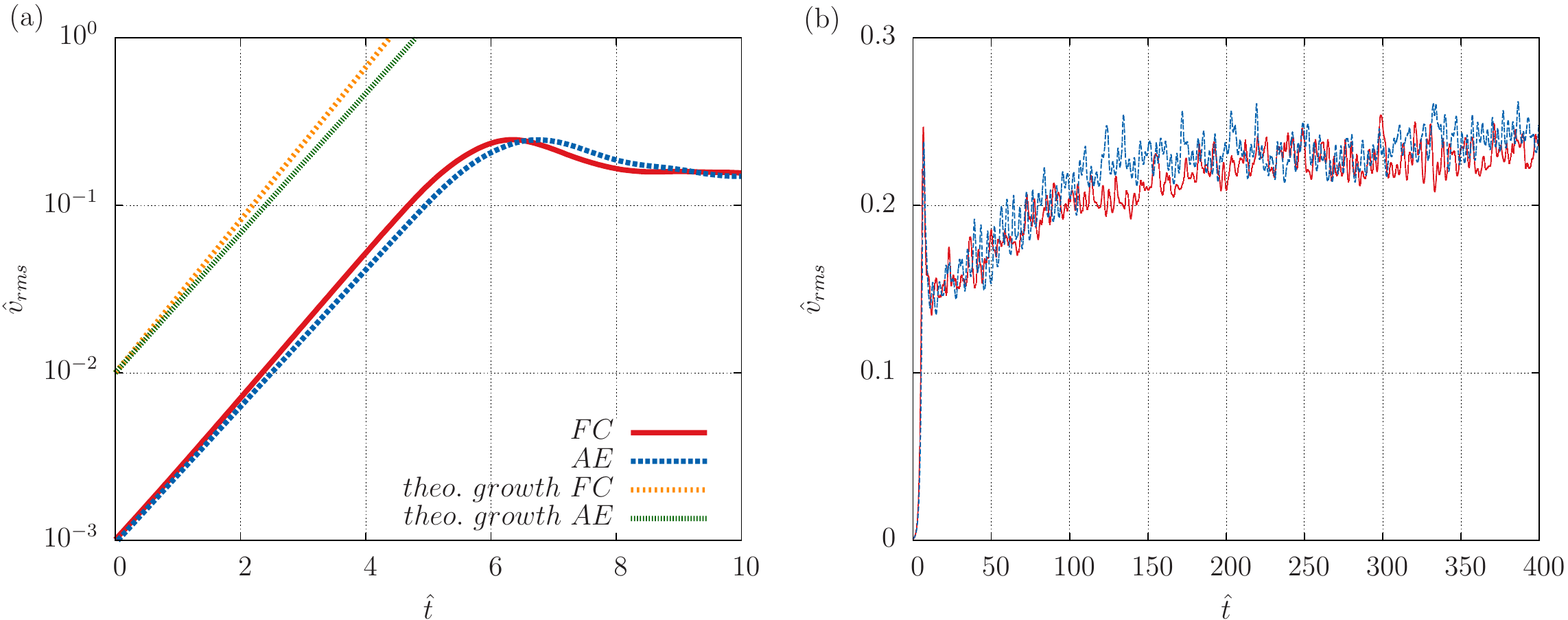}
\caption{The time evolution of the root-mean-square velocities $\nond{v}_{\mathrm {rms}}$ is plotted against time for nonlinear simulations. Panel (a) displays the exponential growth of the convective instability in the nonlinear fully compressible (red) and anelastic (blue) simulations with $Pr=0.1$, $Ek=10^{-6}$ and $Ra=1.98\times10^9$, $N_\rho^{ad}=2$. Both cases match their expected growth rates corresponding to the non-oscillatory instabilities with $\nond{r}_{FC}=1.05$ (orange) and $\nond{r}_{AE}=0.96$ (green). Panel (b) shows the long-term evolution of $v_{\mathrm {rms}}$ for both cases. After the first five free-fall times characterised by exponential growth, a large-scale cyclonic vortex emerges, which accumulates kinetic energy until about $\nond{t}=200$ when the simulations reach statistical equilibrium. The time averaged root-mean-square velocities and their standard deviations for $200\le \nond{t} \le 400$ result in $\left<\nond{v}_{\mathrm {rms}}^{FC}\right> = 0.228 \pm 0.009$ and $\left<\nond{v}_{\mathrm {rms}}^{AE}\right> = 0.233 \pm 0.010$, a difference of about $2\%$. (Colour online)}
\label{figure7}
\end{figure}
Figure~\ref{figure7} shows the time evolution of the root-mean-square velocity
\begin{align}
\nond{v}_{\mathrm {rms}}\, = \,\sqrt{\frac{1}{\nond{V}}\int {\mathrm d}\nond{V}\, \nond{\vec v}^2}\,,
\end{align}
with $\nond{V}$ being the volume of the domain, for the fully compressible and the anelastic simulation in red and blue, respectively. Panel (a) displays the exponential growths of $v_{\mathrm {rms}}$, which match the theoretical predictions for the fully compressible (orange) and anelastic (green) non-oscillatory instabilities.
Panel (b) shows the long-time evolution of $v_{\mathrm {rms}}$ being qualitatively the same for both cases. After the first five free-fall times characterised by exponential growth, one large-scale cyclonic vortex emerges in each simulation, which accumulates kinetic energy until about $\nond{t}=200$ when the computations reach statistical equilibrium.
The time-averaged fully compressible value for $v_{\mathrm {rms}}$ matches the corresponding anelastic one within $2\%$ in this final state. Also the net heat transport in terms of the Nusselt number
\begin{equation}
Nu\,=\,-\, \upartial_z \langle \nond{T}_1 \rangle \bigr|_{\nond{z}=0}\,,
\end{equation}
where the angle brackets imply temporal and horizontal averaging, turns out to be comparable in the final state, with $Nu=13.2$ for the anelastic and $Nu=12.2$ for the fully compressible case representing a difference of roughly $8\%$. This conformity in the output parameters agrees with direct comparisons of anelastic and fully compressible non-rotating convection \citep{Verhoeven2015} and shows (albeit indirectly) that the time derivative of the density fluctuation in the continuity equation is not important in our nonlinear simulations.

\begin{figure}
\setlength{\unitlength}{\linewidth}
\begin{picture}(1.0,0.5)
\thicklines
\put(0.0,0.0){\includegraphics[width=0.5\linewidth]{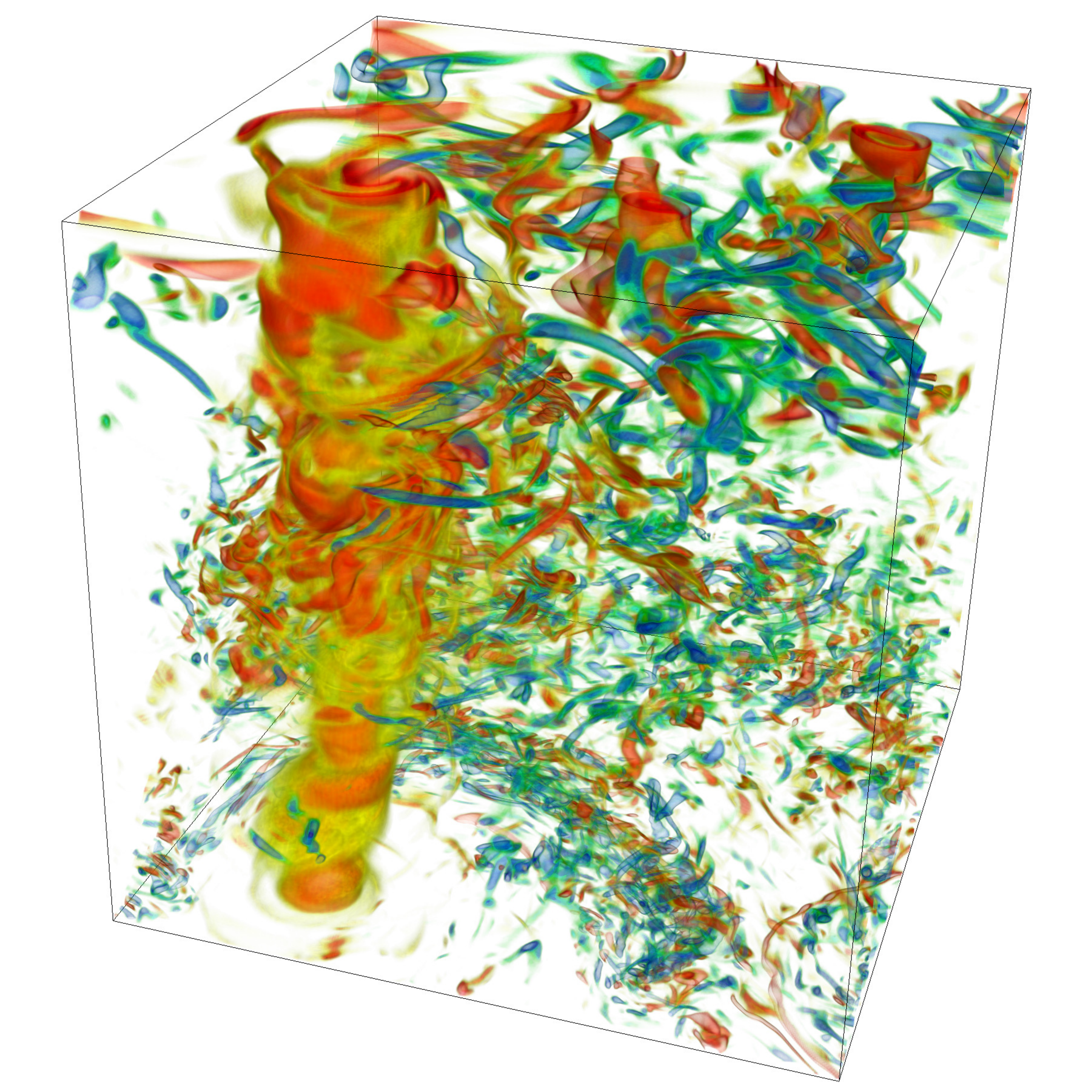}}
\put(0.5,0.0){\includegraphics[width=0.5\linewidth]{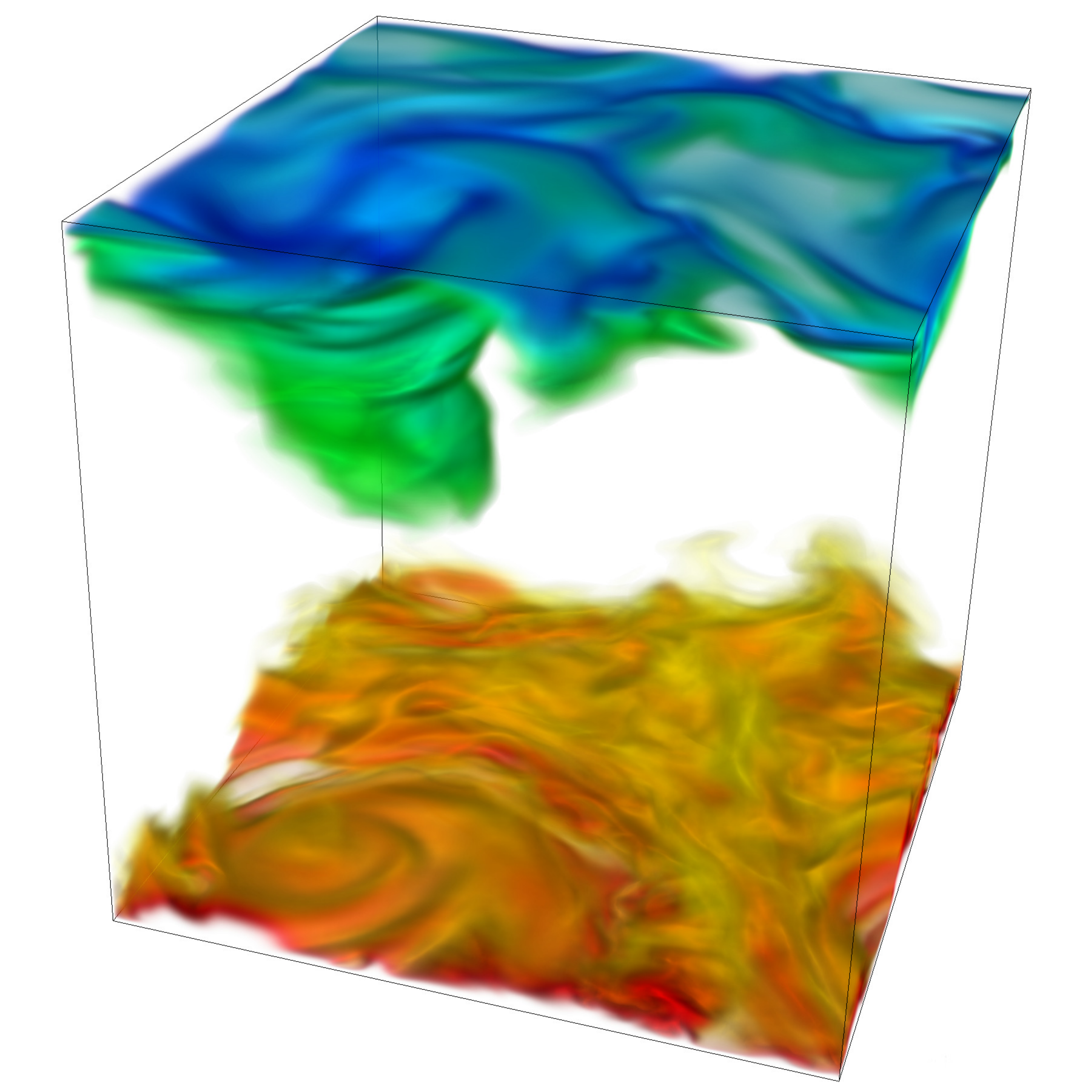}}
\put(0.0,0.47){(a)}
\put(0.5,0.47){(b)}
\end{picture}
\caption{(a) Dynamics in an anelastic simulation run that reached statistical equilibrium, which is illustrated by a volume rendering of the vertical vorticity $\nond{\omega}_z$. While red and yellow colours denote positive $\nond{\omega}_z$, blue and green signify negative $\nond{\omega}_z$. Most conspicuous is the emergence of a large-scale cyclonic vortex, whose angular momentum is balanced by many small-scale anticyclones. (b) The corresponding superadiabatic temperature $\nond{T}_1$ is displayed with red specifying hot and blue indicating cold material. The employed parameters are $Ra=1.98\times10^9$, $Pr=0.1$, $Ek=10^{-6}$ and $N_\rho=2$, which is in the regime of the suspected breakdown of the anelastic approximation. Nevertheless, corresponding snapshots taken from numerical simulations of fully compressible convection look qualitatively the same and are visually indistinguishable from the example at hand. For better visibility, the domain has been horizontally stretched by a factor of $2.5$.}
\label{figure3}
\end{figure}
In order to give a visual impression of the fluid dynamics, figure~\ref{figure3} shows snapshots of the anelastic simulation in statistical equilibrium displaying (a) the vertical vorticity
\begin{align}
\nond{\omega}_z=\nond{\upartial}_x \nond{v}_y - \nond{\upartial}_y \nond{v}_x
\end{align}
and (b) the superadiabatic temperature $\nond{T}_1$. Panel (a) clearly shows the large-scale cyclonic vortex (red), whose angular momentum is balanced by many small-scale anticyclones (blue). In panel (b) cold material (blue) is located near the top and hot fluid (red) close to the bottom boundary.

Such symmetry breaking single large-scale cyclonic vortices typically emerge on the verge of the regime of nonlinear rapid rotation \citep[see, e.g.,][]{Guervilly2014,Favier2014}. Ekman numbers of ${\mathrm O}(10^{-7})$ and below are necessary in order to reach the nonlinear geostrophic turbulence regime, in which strong cyclonic and anticyclonic vortices are generated with the symmetry being restored \citep{Stellmach2014,Rubio2014}. Such parameter ranges are clearly desirable, albeit not accessible with our fully compressible code, which prevents us from carrying out a direct comparison with anelastic results for these extreme parameters. A broader parameter range needs to be tested in future studies.

\section{Conclusions}
\label{conclusions}

The validity of the anelastic approximation in the astrophysically relevant regime of rapidly-rotating compressible convection has recently been questioned by \citet{Calkins2015}. The high computational efficiency and broad application of sound-proof approaches provided the motivation for this paper, that extends their work by further constraining and reviewing the applicability of different sound-proof models in this regime.

As a starting point we focused on the pseudo-incompressible approximation \citep{Durran1989}, which comprises more discreet simplifications in comparison to the anelastic approach.
This is indeed reflected in our computations and in contrast to the anelastic approximation, the pseudo-incompressible approach does not fail in linear stability calculations at the point of marginal stability. Instead its results slowly deviate from the fully compressible solution as the perturbational pressure term in the continuity equation becomes more and more non-negligible, which is usually the case for an increasing number of density scale heights and decreasing Prandtl number shortly after the anelastic approximation breaks down.
Our results confirm what \citet{Calkins2015} conclude about anelastic models of marginally stable convection; i.e., that the anelastic approximation for compressible convection in the rapidly-rotating low Prandtl number regime can be inaccurate at marginal stability.
We find that the anelastic approach breaks down for simulations, in which the sound-crossing time of the computational domain exceeds the rotation time scale. Correspondingly, a rotational Mach number is defined as the ratio of both values, which in analogy to the classical Mach number needs to be small in order for sound-proof approaches to be valid.

As the rotational Mach number is inversely proportional the square-root of the Rayleigh number, the situation is much different for higher supercriticality, where the anelastic approximation neither fails nor gives rise to spurious growth for any parameters investigated. Instead our computed fully compressible growth rates generally converge against the ones of the anelastic limit case as the superadiabaticity decreases.
In this regime, we found the Rayleigh number to be the most constraining parameter in the sense that low values---close to marginal stability---show distinct differences between anelastic and fully compressible results, whereas moderately high values---of the order of the critical value of non-oscillatory convection---were sufficient to yield a match within a few percent for moderately low superadiabaticities.
As Rayleigh numbers in planetary and stellar convection systems are often much higher and superadiabaticities lower than this, we expect the problems causing the breakdown of the anelastic approximation at marginal stability to be insignificant in such systems and predict the agreement between the approximated and the full equations to get even closer.

These findings could be confirmed by fully nonlinear turbulent convection simulations. The computations in the regime where \citet{Calkins2015} suspected the breakdown of the anelastic approximations, showed close qualitative and quantitative agreement between the anelastic and the fully compressible case. However, further work needs to be done in order to exclude effects other than the one suggested by \citet{Calkins2015}, which may cause problems in the anelastic approximation in the geostrophic turbulence regime.

To sum up, the anelastic approximation breaks down for rapidly-rotating compressible convection in low Prandtl number fluids at the point of marginal stability when the rotational Mach number is greater than one. According to our results, however, these problems disappear in the astrophysically more relevant regime of turbulence typically characterised by small rotational Mach numbers.
Although we did not specifically test the pseudo-incompressible approximation for supercritical convection, we do not see any reason why the same should not be true for this less invasive approach.

\section*{Acknowledgments}
We would like to thank two anonymous reviewers for comments that greatly improved the manuscript. We gratefully acknowledge that Prof. Pascale Garaud from UCSC provided her NRK routines, which were the basis of our linear convection code. Support was provided for J.V. by the National Aeronautics and Space Administration under grants OPR NNX13AK94G and PGG NNX14AN70G. The simulations were run at UCSC on the Hyades supercomputer obtained with an MRI grant from the National Science Foundation.

\markboth{J.~Verhoeven and G.A.~Glatzmaier}{Geophysical and Astrophysical Fluid Dynamics}
\bibliographystyle{gGAF}
\bibliography{GGAF-2017-0002_Verhoeven}

\markboth{J.~Verhoeven and G.A.~Glatzmaier}{Geophysical and Astrophysical Fluid Dynamics}

${}$

\newpage

\appendix

\section{Nonlinear equations of compressible convection}
\label{appendix_nonlinear_eq}

Nonlinear equations for fully compressible and anelastic convection for the constant $\mu$ and $k_t$ case as given by \citet{Verhoeven2015} with additionally considering rotation are displayed below. The background profiles $\nond{\rho}_0$, $\nond{T}_0$ and $\nond{p}_0$ are assumed to be adiabatic.

\subsection{Fully compressible equations}
The non-dimensional fully compressible equations read

\begin{subequations}
\begin{align}
\epsilon\nond{\upartial}_{t} \nond{\rho}_{1} + \nond{\vec\nabla}{\bm \cdot}\left[(\nond{\rho}_{0} + \epsilon\nond{\rho}_{1})\nond{\vec v}\right]\,=\, &\,0\,,
\label{continuity_nonl}\\
(\nond{\rho}_{0}+\epsilon\nond{\rho}_{1})\bigl[\nond{\upartial}_{t} \nond{\vec v} + (\nond{\vec v} {\bm \cdot} \nond{\vec\nabla})\nond{\vec v}\bigr]\, =\,&\, -\, \nond{\vec\nabla} \nond{p}_{1} - \nond{\rho}_{1} \hat{\vec z} + \sqrt{\frac{Pr}{Ra}}\bigl[\nond{\vec\nabla}^{2}\nond{\vec v} + \tfrac{1}{3}\nond{\vec\nabla}(\nond{\vec\nabla}{\bm \cdot}\nond{\vec v})\bigr]
\nonumber \\
& \qquad\qquad\qquad - \frac{1}{Ek}\sqrt{\frac{Pr}{Ra}}(\nond{\rho}_0+\epsilon\nond{\rho}_1)\nond{\vec z}\times\nond{\vec v}\,, \label{momentum_nonl}\\
(\nond{\rho}_{0}+\epsilon\nond{\rho}_{1})\bigl[\nond{\upartial}_{t} \nond{T}_{1} + (\nond{\vec v} {\bm \cdot} \nond{\vec\nabla}) \nond{T}_{1}\bigr]\hskip 6mm &\nonumber \\
-D \nond{\rho}_{1} \nond{v}_{z}  - D \bigl[\nond{\upartial}_{t} \nond{p}_{1} + (\nond{\vec v} {\bm \cdot} \nond{\vec\nabla}) \nond{p}_{1}\bigr]\, =\, &\,\frac{1}{\sqrt{Ra Pr}} \nond{\vec\nabla}^{2}\nond{T}_1 + 2D\sqrt{\frac{Pr}{Ra}} \bigl[\nond{e}_{ij} - \tfrac{1}{3}(\nond{\vec\nabla}{\bm \cdot}\nond{\vec v})\delta_{ij}\bigr]^{2}, \qquad\label{energy_nonl}\\
\frac{D}{1-({1}/{\gamma})}\frac{\nond{p}_{1}}{\nond{p}_{0}}\, =\, &\,\frac{\nond{T}_{1}}{\nond{T}_{0}} + \frac{\nond{\rho}_{1}}{\nond{\rho}_{0}} + \epsilon \frac{\nond{\rho}_{1}}{\nond{\rho}_{0}} \frac{\nond{T}_{1}}{\nond{T}_{0}}\,,
\label{state_nonl}
\end{align}
\end{subequations}
%
%
with $\nond{e}_{ij}=\frac{1}{2}\bigl(\nond{\upartial}_{j} \nond{v}_{i} + \nond{\upartial}_{i} \nond{v}_{j}\bigr)$ being the strain rate tensor.

\subsection{Anelastic equations}
The non-dimensional anelastic equations result from (\ref{continuity_nonl}-d)
in the limit of $\epsilon\rightarrow 0$:
\begin{subequations}
\begin{align}
\nond{\vec\nabla}{\bm \cdot}\left(\nond{\rho}_{0}\nond{\vec v}\right)\,=\,&\,0\,, \label{continuity_nonl_ae}\\
\nond{\rho}_{0}\bigl[\nond{\upartial}_{t} \nond{\vec v} + (\nond{\vec v} {\bm \cdot} \nond{\vec\nabla})\nond{\vec v}\bigr]\, = \,&\,- \nond{\vec\nabla} \nond{p}_{1} - \nond{\rho}_{1} \hat{\vec z} + \sqrt{\frac{Pr}{Ra}}\bigl[\nond{\vec\nabla}^{2}\nond{\vec v} + \tfrac{1}{3}\nond{\vec\nabla}(\nond{\vec\nabla}{\bm \cdot}\nond{\vec v})\bigr]\nonumber\\
&\qquad\qquad\qquad - \frac{1}{Ek}\sqrt{\frac{Pr}{Ra}}\nond{\rho}_0\nond{\vec z}\times\nond{\vec v}\,, \label{momentum_nonl_ae}\\
\nond{\rho}_{0}\bigl[\nond{\upartial}_{t} \nond{T}_{1} + (\nond{\vec v} {\bm \cdot} \nond{\vec\nabla}) \nond{T}_{1}\bigr]\hskip 6mm &\nonumber\\
-D \nond{\rho}_{1} \nond{v}_{z} - D \bigl[\nond{\upartial}_{t} \nond{p}_{1} + (\nond{\vec v} {\bm \cdot} \nond{\vec\nabla}) \nond{p}_{1}\bigr]\, =\,&\, \frac{1}{\sqrt{Ra Pr}} \nond{\vec\nabla}^{2}\nond{T}_1 + 2D\sqrt{\frac{Pr}{Ra}} \bigl[\nond{e}_{ij} - \tfrac{1}{3}(\nond{\vec\nabla}{\bm \cdot}\nond{\vec v})\delta_{ij}\bigr]^{2}, \qquad\label{energy_nonl_ae}\\
\frac{D}{1-({1}/{\gamma})}\frac{\nond{p}_{1}}{\nond{p}_{0}} \,= \,&\,\frac{\nond{T}_{1}}{\nond{T}_{0}} + \frac{\nond{\rho}_{1}}{\nond{\rho}_{0}}\,.\label{state_nonl_ae}
\end{align}
\end{subequations}
%

\section{Mach number}
\label{appendix_mach_reg}

The Mach number $M$ is defined as the ratio of a reference velocity, which we assume to be the free-fall velocity
\be
v_{r} \, = \,\sqrt{\frac{gd\Delta\rho}{\rho_r}}
\ee
to the sound speed
\be
v_{\mathrm {sound}} \,= \,\sqrt{\frac{c_p(c_p-c_v)T}{c_v}}\,=\, \sqrt{\frac{c_p(c_p-c_v)T_r}{c_v}} \,\nond{\rho}_0^{1/(2n)}
\ee
resulting in
\be
M\, =\,\frac{v_r}{v_{\mathrm {sound}}}\, =\,\sqrt{\frac{c_v g d \Delta\rho}{c_p(c_p-c_v)T_r \rho_r}} \,\nond{\rho}_0^{-1/(2n)}
\,=\,\sqrt{\frac{\epsilon D}{\gamma - 1}}\, \nond{\rho}_0^{-1/(2n)}.\label{machnumber}
\ee
Exploiting $\nond{\rho}_0^{\mathrm {top}}=\exp{\left(-N_\rho\right)}$ yields the Mach number at the top boundary,
\be
\label{Mach_top}
M^{\mathrm {top}}\,= \,\sqrt{\frac{\epsilon D}{\gamma - 1}}\, \exp{\left(\frac{N_\rho}{2n}\right)}\,,
\ee
where the sound speed is the lowest and thus $M$ has a maximum.

\section{Phase Mach number}
\label{appendix_mach_phase}

The phase Mach number $M_{\mathrm {phase}}$ will be defined as the ratio of the phase velocity corresponding to the oscillatory instability
\be
v_{\mathrm {phase}} \,=\, \frac{\omega}{k}\, =\, \sqrt{\frac{g d \Delta T}{T_r}}\, \frac{\nond{\omega}}{\nond{k}}
\ee
to the sound speed
\be
v_{\mathrm {sound}}\, =\, \sqrt{\frac{c_p(c_p-c_v)T}{c_v}} \,=\, \sqrt{\frac{c_p(c_p-c_v)T_r}{c_v}} \,\nond{\rho}_0^{1/(2n)}
\ee
resulting in
\begin{align}
M_{\mathrm {phase}} \,=\,\frac{v_{\mathrm {phase}}}{v_{\mathrm {sound}}}\,=\,&\,\sqrt{\frac{c_v g d \Delta T}{c_p(c_p-c_v)T_r^2}}\, \frac{\nond{\omega}}{\nond{k}} \,\nond{\rho}_0^{-1/(2n)}\nonumber\\
=\,&\,\sqrt{\frac{g d}{c_pT_r}}\sqrt{\frac{c_v}{c_p-c_v}}\sqrt{\frac{\Delta T}{T_r}}\, \frac{\nond{\omega}}{\nond{k}}\, \nond{\rho}_0^{-1/(2n)}\,= \,\sqrt{\frac{\epsilon D}{\gamma - 1}}\, \frac{\nond{\omega}}{\nond{k}}\, \nond{\rho}_0^{-1/(2n)}\,.
\end{align}
Exploiting $\nond{\rho}_0^{\mathrm {top}}=\exp{\left(-N_\rho\right)}$ and equation (\ref{Mach_top}) yields the phase Mach number at the top boundary,
\be
M_{\mathrm {phase}}^{\mathrm {top}}\,= \,\sqrt{\frac{\epsilon D}{\gamma - 1}}\, \frac{\nond{\omega}}{\nond{k}}\, \exp{\left(\frac{N_\rho}{2n}\right)}\,= \,M^{\mathrm {top}} \,\frac{\nond{\omega}}{\nond{k}}\,,
\ee
where the sound speed is the lowest and thus $M_{\mathrm {phase}}$ has a maximum.

\section{Rotational Mach number}
\label{appendix_mach_rot}

The rotational Mach number $M_{\mathrm {rot}}$ will be defined as the ratio of the velocity corresponding to the domain depth $d$ and the angular frequency $\Omega$, where
\be
v_{\mathrm {rot}} \,= \,2\Omega d\,,
\ee
to the sound speed at the bottom
\be
v_{\mathrm {sound}}^{\mathrm {bot}} \,= \,\sqrt{\frac{c_p(c_p-c_v)T_r}{c_v}}
\ee
resulting in
\be
M_{\mathrm {rot}} \,=\,\frac{v_{\mathrm {rot}}}{v_{\mathrm {sound}}^{\mathrm {bot}}}\,=\,\sqrt{\frac{4 c_v \Omega^2 d^2}{c_p(c_p-c_v)T_r}}\,= \,\frac{1}{Ek}\sqrt{\frac{\epsilon D Pr}{(\gamma - 1) Ra}}\,.
\ee
This is equivalent to defining the rotational Mach number as the ratio of the sound-crossing time of the domain to a rotation time scale,
\be
M_{\mathrm {rot}} \,= \,\frac{t^{\mathrm {bot}}_{\mathrm {sound}}}{t_{\mathrm {rot}}}\,=  \,\frac{2\Omega d}{v^{\mathrm {bot}}_{\mathrm {sound}}} \,,
\ee
based on the sound speed $v^{\mathrm {bot}}_{\mathrm {sound}}$ at the bottom boundary. Note that the value of $t^{\mathrm {bot}}_{\mathrm {sound}}$ is similar to the one of the real sound-crossing time $t_{\mathrm {sound}}=\int_0^d {\mathrm d}z / v_{\mathrm {sound}}$.

\section{Numerically solved linear equations}
\label{appendix_numerics}
This section summarises the equations as they are solved numerically in the linear convection code, which is based on a Newton-Raphson-Kantorovich (NRK) method. As it only deals with non-dimensional quantities, the hat $\nond{}$ is left out for clarity. Equations (\ref{continuity_momentum_heat_state_therm-relation_nondim1}a-e)
can be simplified by expressing $\rho_1$ and $s_1$ in terms of ${T}_1$ and ${p}_1$, which reduces the number of equations from five to three,
\begin{subequations}
\begin{gather}
\label{continuity_nondim2}
\alpha \frac{\epsilon D}{\gamma - 1}\frac{1}{{T}_0}{\upartial}_t {p}_1 + \beta \left(\epsilon D\frac{{\rho}_0}{{p}_0}{\upartial}_t {p}_1 - \epsilon \frac{{\rho}_0}{{T}_0} {\upartial}_t {T}_1\right) \,=\, - \,{\vec\nabla}{\bm \cdot}\left({\rho}_0 {\vec v}\right), \\
\label{momentum_nondim2}
{\rho}_0{\upartial}_t {\vec v}\, =\, -\, {\vec\nabla} {p}_1 + \left(\frac{{\rho}_0}{{T}_0} {T}_1 - \frac{\gamma}{\gamma - 1} D \frac{{\rho}_0}{{p}_0}{p}_1\right) {\vec z} + \sqrt{\frac{Pr}{Ra}}{\vec\nabla}{\bm \cdot}{\tensor\Pi} - \frac{1}{Ek}\sqrt{\frac{Pr}{Ra}}{\rho}_0{\vec z}\times{\vec v}, \qquad\\
\label{heat_nondim2}
{\rho}_0 {\upartial}_t {T}_1  - D {\upartial}_t {p}_1\,=\, \rho_0 v_z + \frac{1}{\sqrt{Pr Ra}}{\vec\nabla}\left({k}_t {\vec\nabla} {T}_1\right).
\end{gather}
\end{subequations}
The parameters $\alpha$ and $\beta$ have been introduced in order to account for the three different approaches under investigation. While in the anelastic approximation the perturbational density term is neglected in the continuity equation (i.e., $\alpha = \beta = 0$), the pseudo-incompressible approximation just neglects the perturbational pressure term resulting in $\alpha = 0$ and $\beta = 1$.  The fully compressible equations do not contain any simplifications, thus $\alpha = \beta = 1$. We further define $P_1 = \sqrt{Ra/Pr}\, p_1$ in order to simplify the numerical method, compare equation (\ref{momentum_nondim2}) with (\ref{dz_vz_re_equation}), (\ref{dz_vz_im_equation}).

For the constant dynamic viscosity $\mu$ and thermal conductivity $k_t$ case we assume
\bme
\begin{align}
\mu(z) \,=\, &\, 1\,, &    \mu'(z)\, =\, &\, 0\,, \\
  k_t(z) \,= \,&\, 1\,, &   k_t'(z)\, =\, &\, 0\,,\qquad\qquad\qquad\qquad
\end{align}
\eme
whereas for constant diffusivities $\nu$ and $\kappa$ the corresponding dynamic viscosity and thermal conductivity vary with depth (see (\ref{viscosity-conductivity}a,b) and (\ref{T_0-rho_0-p_0-s_0}b)):
\bme
\begin{align}
\mu(z)\, =\, & \,T_0^n(z)\,, &  \mu'(z) \,=\, &\, - \left(D + \epsilon\right) n T_0^{n - 1}(z)\,, \\
\qquad k_t(z)\, =\, &\, T_0^n(z) \,, & k_t'(z) \,=\, &\, - \left(D + \epsilon\right) n T_0^{n - 1}(z)\,.
\end{align}
\eme
Note that the $'$ sign denotes first derivatives with respect to $z$, i.e., the numerically independent variable $\mu'(z)$ is the analytical derivative of $\mu(z)$.

When using the typical normal mode ansatz, e.g., ${T}_1={T}({z})\exp\left[{r}{t}+{\mathrm i}\left({\omega} {t} + {k}_x {x} + {k}_y {y}\right)\right]$, with the wavenumber $k_x$ and $k_y$ not to be confused with the thermal conductivity $k_t$, equations (\ref{continuity_nondim2}-c)
can be solved separately for the real part (index $_{\mathrm{re}}$),
\begin{subequations}
\begin{align}
\frac{\mathrm d}{{\mathrm d} z} {v}_{x,\mathrm{re}} \,= \, & \, v_{x,\mathrm{re}}'\,,\\[0.1em]
\mu \frac{\mathrm d}{{\mathrm d} z} v_{x,\mathrm{re}}' \,=\, & \,\sqrt{\frac{Ra}{Pr}}r T_0^n v_{x,\mathrm{re}} - \sqrt{\frac{Ra}{Pr}}\omega T_0^n v_{x,\mathrm{im}} - k_x P_{im} + \mu\left(\tfrac{4}{3}k_x^2+k_y^2\right) v_{x,\mathrm{re}} \nonumber\\
& + \tfrac{1}{3}\mu k_x k_y v_{y,\mathrm{re}} + \tfrac{1}{3}\mu k_x v_{z,\mathrm{im}}' - \mu' v_{x,\mathrm{re}}' + \mu' k_x v_{z,\mathrm{im}} - \frac{1}{Ek}T_0^n v_{y,\mathrm{re}}\,,\qquad\qquad\\[0.2em]
\frac{\mathrm d}{{\mathrm d} z} v_{y,\mathrm{re}}\, =\, & \,v_{y,\mathrm{re}}'\,,\\[0.1em]
\mu \frac{\mathrm d}{{\mathrm d} z} v_{y,\mathrm{re}}'\, =\, &\, \sqrt{\frac{Ra}{Pr}}r T_0^n v_{y,\mathrm{re}} - \sqrt{\frac{Ra}{Pr}}\omega T_0^n v_{y,\mathrm{im}} - k_y P_{im} + \mu\left(k_x^2+\tfrac{4}{3}k_y^2\right)v_{y,\mathrm{re}} \nonumber\\
& + \tfrac{1}{3} \mu k_x k_y v_{x,\mathrm{re}} + \tfrac{1}{3}\mu k_y v_{z,\mathrm{im}}' - \mu' v_{y,\mathrm{re}}' + \mu' k_y v_{z,\mathrm{im}} + \frac{1}{Ek} T_0^n v_{x,\mathrm{re}}\,,\\[0.2em]
\frac{\mathrm d}{{\mathrm d} z} v_{z,\mathrm{re}}\, =\, &\, v_{z,\mathrm{re}}'\,,\\[0.1em]
\tfrac{4}{3} \mu \frac{\mathrm d}{{\mathrm d} z} v_{z,\mathrm{re}}' \hskip 5mm &\nonumber\\
- \frac{\mathrm d}{{\mathrm d} z} P_{\mathrm{re}}\,=\, & \,\sqrt{\frac{Ra}{Pr}}r T_0^n v_{z,\mathrm{re}} - \sqrt{\frac{Ra}{Pr}}\omega T_0^n v_{z,\mathrm{im}} + D (n_{ad}+1)\frac{1}{T_0} P_{\mathrm{re}} \nonumber\\
&\, - \sqrt{\frac{Ra}{Pr}} T_0^{n-1} T_{\mathrm{re}} + \mu\left(k_x^2+k_y^2\right) v_{z,\mathrm{re}}\nonumber \\
&\,+ \tfrac{1}{3}\mu k_x v_{x,\mathrm{im}}' + \tfrac{1}{3}\mu k_y v_{y,\mathrm{im}}' - \tfrac{4}{3}\mu' v_{z,\mathrm{re}}' - \tfrac{2}{3}\mu' k_x v_{x,\mathrm{im}} - \tfrac{2}{3}\mu' k_y v_{y,\mathrm{im}}\,,\qquad\label{dz_vz_re_equation}\\[0.2em]
\frac{\mathrm d}{{\mathrm d} z} T_{\mathrm{re}}\, =\, & \,T_{\mathrm{re}}'\,,\\[0.1em]
k_t \frac{\mathrm d}{{\mathrm d} z} T_{\mathrm{re}}'\, =\, &\, \sqrt{Pr Ra}\, r T_0^n T_{\mathrm{re}} - \sqrt{Pr Ra}\, \omega T_0^n T_{\mathrm{im}}  - D Pr\, r P_{\mathrm{re}} + D Pr\, \omega P_{\mathrm{im}}\nonumber \\
& \,+ k_t \left(k_x^2+k_y^2\right) T_{\mathrm{re}} - k_t' T_{\mathrm{re}}' - \sqrt{Pr Ra} \,T_0^n v_{z,\mathrm{re}}\,, \\[0.2em]
0 \,=\, &\, \left(\alpha n_{ad} + \beta \right) \epsilon D \sqrt{Pr}\, r \frac{1}{T_0^{n+1}} P_{\mathrm{re}} - \left(\alpha n_{ad} + \beta \right) \epsilon D \sqrt{Pr}\, \omega \frac{1}{T_0^{n+1}} P_{\mathrm{im}} \nonumber\\
&\, - \beta \epsilon \sqrt{Ra} \,r  \frac{1}{T_0} T_{\mathrm{re}} + \beta \epsilon \sqrt{Ra}\, \omega  \frac{1}{T_0} T_{\mathrm{im}} - n\left(D+\epsilon\right)\sqrt{Ra}\frac{1}{T_0} v_{z,\mathrm{re}} \nonumber\\
& \,+ \sqrt{Ra\,}v_{z,\mathrm{re}}' - \sqrt{Ra}\,k_x v_{x,\mathrm{im}} - \sqrt{Ra}\,k_y v_{y,\mathrm{im}}\,,
\end{align}
\end{subequations}
and the imaginary part (index $_{\mathrm{im}}$)
\begin{subequations}
\begin{align}
\frac{\mathrm d}{{\mathrm d} z} v_{x,\mathrm{im}}\, =\, &\, v_{x,\mathrm{im}}'\,,\\[0.1em]
\mu \frac{\mathrm d}{{\mathrm d} z} v_{x,\mathrm{im}}'\, = \,&\, \sqrt{\frac{Ra}{Pr}}r T_0^n v_{x,\mathrm{im}} + \sqrt{\frac{Ra}{Pr}}\omega T_0^n v_{x,\mathrm{re}} + k_x P_{\mathrm{re}} + \mu\left(\tfrac{4}{3}k_x^2+k_y^2\right) v_{x,\mathrm{im}} \nonumber\\
&\, + \tfrac{1}{3}\mu k_x k_y v_{y,\mathrm{im}} - \tfrac{1}{3}\mu k_x v_{z,\mathrm{re}}'  - \mu' v_{x,\mathrm{im}}' - \mu' k_x v_{z,\mathrm{re}} - \frac{1}{Ek}T_0^n v_{y,\mathrm{im}}\,,\qquad\quad\\[0.2em]
\frac{\mathrm d}{{\mathrm d} z} v_{y,\mathrm{im}} \,=\, &\, v_{y,\mathrm{im}}'\,,\\[0.1em]
\mu \frac{\mathrm d}{{\mathrm d} z} v_{y,\mathrm{im}}'\, =\, & \, \sqrt{\frac{Ra}{Pr}}r T_0^n v_{y,\mathrm{im}} + \sqrt{\frac{Ra}{Pr}}\omega T_0^n v_{y,\mathrm{re}} + k_y P_{\mathrm{re}} + \mu\left(k_x^2+\tfrac{4}{3}k_y^2\right) v_{y,\mathrm{im}}\nonumber \\
&\, + \tfrac{1}{3} \mu k_x k_y v_{x,\mathrm{im}} - \tfrac{1}{3}\mu k_y v_{z,\mathrm{re}}'  - \mu' v_{y,\mathrm{im}}' - \mu' k_y v_{z,\mathrm{re}} + \frac{1}{Ek}T_0^n v_{x,\mathrm{im}}\,,\\[0.2em]
\frac{\mathrm d}{{\mathrm d} z} v_{z,\mathrm{im}}\, =\, &\, v_{z,\mathrm{im}}'\,,\\[0.1em]
\tfrac{4}{3}\mu\frac{\mathrm d}{{\mathrm d} z} v_{z,\mathrm{im}}' \hskip 5mm &\nonumber\\
- \frac{\mathrm d}{{\mathrm d} z} P_{\mathrm{im}} \,=\, &\, \sqrt{\frac{Ra}{Pr}}r T_0^n v_{z,\mathrm{im}} + \sqrt{\frac{Ra}{Pr}}\omega T_0^n v_{z,\mathrm{re}} + D\left(n_{ad}+1\right)\frac{1}{T_0} P_{\mathrm{im}}\nonumber \\
&\,- \sqrt{\frac{Ra}{Pr}}T_0^{n-1} T_{\mathrm{im}} + \mu\left(k_x^2+k_y^2\right) v_{z,\mathrm{im}} \nonumber\\
&\, - \tfrac{1}{3}\mu k_x v_{x,\mathrm{re}}' - \tfrac{1}{3}\mu k_y v_{y,\mathrm{re}}' - \tfrac{4}{3}\mu' v_{z,\mathrm{im}}' + \tfrac{2}{3}\mu' k_x v_{x,\mathrm{re}} + \tfrac{2}{3}\mu' k_y v_{y,\mathrm{re}}\,,\label{dz_vz_im_equation}\\
\frac{\mathrm d}{{\mathrm d} z} T_{\mathrm{im}} \,= \,&\, T_{\mathrm{im}}'\,,\\[0.1em]
k_t \frac{\mathrm d}{{\mathrm d} z} T_{\mathrm{im}}' \,=\, &\, \sqrt{Pr Ra}\,r T_0^n T_{\mathrm{im}} + \sqrt{Pr Ra}\,\omega T_0^n T_{\mathrm{re}} - D Pr\, r P_{\mathrm{im}} - D Pr \,\omega P_{\mathrm{re}}\nonumber \\
&  + k_t\left(k_x^2+k_y^2\right) T_{\mathrm{im}} - k_t' T_{\mathrm{im}}' - \sqrt{Pr Ra}\, T_0^n v_{z,\mathrm{im}}\,,\\[0.2em]
0\, =\, &\, \left(\alpha n_{ad} + \beta \right) \epsilon D \sqrt{Pr} r \frac{1}{T_0^{n+1}} P_{\mathrm{im}} + \left(\alpha n_{ad} + \beta \right) \epsilon D \sqrt{Pr} \,\omega \frac{1}{T_0^{n+1}} P_{\mathrm{re}}\nonumber \\
&\, - \beta \epsilon \sqrt{Ra}\, r \frac{1}{T_0} T_{\mathrm{im}} - \beta \epsilon \sqrt{Ra}\, \omega \frac{1}{T_0} T_{\mathrm{re}} - n\left(D+\epsilon\right)\sqrt{Ra}\frac{1}{T_0} v_{z,\mathrm{im}} \nonumber\\
&\, + \sqrt{Ra}\,v_{z,\mathrm{im}}' + \sqrt{Ra}\,k_x v_{x,\mathrm{re}} + \sqrt{Ra}\,k_y v_{y,\mathrm{re}}\,.
\end{align}
\end{subequations}
The equations for the eigenvalues
\bme
\be
\te
\frac{\mathrm d}{{\mathrm d} z} \sqrt{Ra} \,=  \,0\,, \qquad\qquad
\frac{\mathrm d}{{\mathrm d} z} r\, =\,  0\,, \qquad\qquad
\frac{\mathrm d}{{\mathrm d} z} \omega\, = \, 0
\ee
\eme
guarantee that they are constants, independent of $z$.

\end{document}